\documentclass[showpacs,prl,onecolumn,aps,superscriptaddress,preprintnumbers,nofootinbib]{revtex4}
\usepackage[T1]{fontenc}
\usepackage[latin9]{inputenc}
\usepackage{amsmath,amssymb}
\usepackage{epsfig}
\usepackage{graphicx}
\usepackage{amsmath}
\usepackage{amsfonts}
\usepackage{subfig}
\usepackage{epstopdf}
\def\slashchar#1{\setbox0=\hbox{$#1$}     		
   \dimen0=\wd0                                 	
   \setbox1=\hbox{/} \dimen1=\wd1               	
   \ifdim\dimen0>\dimen1                        	
      \rlap{\hbox to \dimen0{\hfil/\hfil}}      	
      #1                                        	
   \else                                        	
      \rlap{\hbox to \dimen1{\hfil$#1$\hfil}}   	
      /                                         	
   \fi}

\renewcommand{\vec}{\boldsymbol}
\newcommand{\beq}{\begin{equation}}
\newcommand{\eeq}{\end{equation}}
\newcommand{\bea}{\begin{eqnarray}}
\newcommand{\eea}{\end{eqnarray}}
\newcommand{\ba}{\begin{array}}
\newcommand{\ea}{\end{array}}

\def\eq#1{{Eq.~(\ref{#1})}}
\def\fig#1{{Fig.~\ref{#1}}}
\newcommand{\bas}{\bar{\alpha}_S}
\newcommand{\as}{\alpha_S}
\newcommand{\nn}{\nonumber}

\newcommand{\h}{\frac{1}{2}}

\newcommand{\Lb}{\left(}
\newcommand{\Rb}{\right)}
\def\pom{{I\!\!P}}

\begin{document}

\title{Bose-Einstein correlations in perturbative QCD: $v_n$ dependence on multiplicity.}
\author{E. ~Gotsman}
\email{gotsman@post.tau.ac.il}
\affiliation{Department of Particle Physics, School of Physics and Astronomy,
Raymond and Beverly Sackler
 Faculty of Exact Science, Tel Aviv University, Tel Aviv, 69978, Israel}
 \author{ E.~ Levin}
\email{leving@post.tau.ac.il, eugeny.levin@usm.cl}
\affiliation{Department of Particle Physics, School of Physics and Astronomy,
Raymond and Beverly Sackler
 Faculty of Exact Science, Tel Aviv University, Tel Aviv, 69978, Israel}
 \affiliation{Departemento de F\'isica, Universidad T\'ecnica Federico
 Santa Mar\'ia, and Centro Cient\'ifico-\\
Tecnol\'ogico de Valpara\'iso, Avda. Espana 1680, Casilla 110-V,
 Valpara\'iso, Chile} 
 
\date{\today}


\begin{abstract}
In this paper we study the dependence of  Bose-Einstein correlations on
 the multiplicity of an event.
 We found that   events with large multiplicity, stem from  the 
production of
 several parton showers, while the additional production of small multiplicity
  in the central rapidity region  (central diffraction),
gives a
 negligible contribution due to emission of soft gluons,
 that leads to the 
Sudakov suppression of the exclusive production of two gluon jets.
 Hence, the Bose-Einstein correlation is the main source of the
 azimuthal angle correlations which generates $v_n$ with odd and even $n$.
  We found, that without this suppression, the measurement of an event with
 given multipilicity, yields $v_{n,n}  < 0$ for odd $n$.
 It appears 
that 
in hadron-nucleus and nucleus-nucleus collisions, the Bose-Einstein
 correlations do not depend on 
 multiplicity, while for hadron-hadron scattering, such dependence
 can be considerable.
 We proposed a simple Kharzeev-Levin-Nardi (KLN) type model, to describe 
the dependence of azimuthal angle correlations
 on the centrality of the event,  in ion-ion collisions.
 
  \end{abstract}
 
 \preprint{TAUP-3007/16}

\maketitle

\tableofcontents

\section{Introduction}
In this paper we continue to discuss the Bose-Einstein correlations
 of gluons as  being the main source of 
the strong azimuthal angle ($ \varphi$)   correlations, that have been
 observed experimentally, in
 nucleus-nucleus, hadron-nucleus and hadron hadron collisions 
\cite{CMSPP,STARAA,PHOBOSAA,STARAA1,CMSPA,CMSAA,ALICEAA,ALICEPA,
ATLASPP,ATLASPA,ATLASAA}. It has been known for  some time in
 the framework of Gribov Pomeron 
Calculus, that the 
Bose-Einstein correlations which stem from the exchange of two
 Pomerons lead to  azimuthal angle
 correlations\cite{PION} (see also Ref.\cite{GLMBEH}),  
 which do not depend on the rapidity
 difference between measured hadrons ( large range rapidity
 (LRR) correlations).  In the framework of QCD, these azimuthal
 correlations  originate  from the production of two patron showers,
 and have been re-discovered in Refs.\cite{KOLU1,KOWE,KOLUCOR,GOLE,KOLULAST} (see
 also Ref.\cite{RAJUREV,KOLUREV}). 
 In
 Ref.\cite{GLMBE} it  was demonstrated 
that  Bose-Einstein correlations  generate $v_n$ with even and odd $n$,
 with  values  which are close to the experimental observed ones.

The goal of this paper is to answer three questions: (i) Is the symmetry
 $\varphi \,\to \pi - 
\varphi$
an inherent property of QCD, or of the colour glass condensate
 (CGC) approach, which is the effective theory
 of QCD at high energies, or it is based on the model assumptions ?
 (ii) What is the multiplicity
 dependence of the azimuthal angle correlations which stem from
 the Bose-Einstein ones?  (iii)
 Is it possible to build a simple KLN-type
 \cite{KLN0,KLN1,KLN2,KLN3,KLN4,KLN5,KLN6} approach to
 describe  azimuthal correlations in nucleus-nucleus collisions ?

 The following are our answers to these questions : 
The symmetry
 $\varphi \,\to\,\pi - 
\varphi$, is not a general
 feature of the QCD (or CGC)
 approach. It  does not stem from the Bose-Einstein correlations of
 identical gluons,
and can only appear  in  measurements that mix  events with different
 multiplicities.
 In the case of hadron-hadron collisions, for example, such symmetry exists
 in the Born approximation of perturbative QCD, and
 could  only  be measured, if
 experimentally the central diffraction production and  the event with
 double multiplicity
 ( $n = 2 \bar n$, where $\bar n$ is the average multiplicity in inclusive
 production) are
 measured and  summed. 
However, 
 the emission of soft gluons for the central exclusive production in 
the Double
 Log Approximation of perturbative QCD, leads to a Sudakov form factor 
which
  suppress this contribution.  Therefore,  the Bose-Einstein correlations
 prevail, leading to $v_n \,\neq\,0$ for odd $n$, even in totally 
inclusive
 measurements, without selection of an event with given multiplicities.

We expect a very mild dependence of $v_n$ on the multiplicity   of the
 observed events. We suggest a
  model for the Bose-Einstein correlations in heavy ion collisions in
 the spirit of the KLN approach,
 which is based on the  concept of constructing  the simplest 
model that takes
into account the discussed phenomena: in our case, the saturation of
 the gluon density and the Bose-Einstein correlations.

The double inclusive cross section of two identical gluons has the
 following general form:
\beq \label{I1}
 \frac{d^2 \sigma}{d y_1 \,d y_2  d^2 p_{T1} d^2 p_{T2}}\Lb \rm identical\,\,
 gluons\Rb\,\,=\,\,
  \, \frac{d^2 \sigma}{d y_1 \,d y_2  d^2 p_{T1} d^2 p_{T2}}\Lb \rm different
 \,\, gluons\Rb \Big( 1 \,+\,C\Lb L_c |\vec{p}_{T2} -   \vec{p}_{T1}|\Rb\Big)
  \eeq
where $C\Lb L_c |\vec{p}_{T2} -   \vec{p}_{T1}|\Rb$  denotes the 
correlation
 function and $L_c$   
 the correlation length.
\eq{I1} is in accord with  Hanbury Brown and Twiss formula (see Refs.
 \cite{HBT,IPCOR})
  \beq \label{I2}
  \frac{d^2 \sigma}{d y_1 \,d y_2  d^2 p_{T1} d^2 p_{T2}}\Lb \rm identical\,\,
 gluons\Rb\,\,\propto\,\,\Big{ \langle}  1\,\,+\,\,e^{i
 r_\mu Q_\mu}\Big{\rangle}
 \eeq
 where  averaging $\langle \dots \rangle$ includes the integration
 over $r_\mu = r_{1,\mu} - r_{2,\mu}$.  There  is only one difference:  
  $Q_\mu = p_{1,\mu} \,-\,p_{2,\mu}$ degenerates to $\vec{Q}\,\equiv
 \,\vec{p}_{T,12} = \vec{p}_{T2} -   \vec{p}_{T1}$,  as
 the production of two gluons
 from the two parton showers, does not depend on rapidities. 
 
   \eq{I2} allows us to measure the typical $r_\mu$
 of the interaction, or in other words, $L_c$ in \eq{I1} is determined by
 the typical volume of the 
interaction. Therefore, we expect several typical $L_c$: the size of  the
 nucleus $R_A$; the nucleon
 size $R_N$ and 
  the typical size, related to the saturation scale
 ($r_{\rm sat}\,=1/Q_s$, where $Q_s$  denotes the saturation 
scale\cite{KOLEB}). Indication of all these sizes have been seen in Bose -
 Einstein correlations
 (see Ref.\cite{GLMBE,GOLE}).
  Using 
\eq{I1}
 we can find $v_n$, since
   \beq \label{I3}
     \frac{d^2 \sigma}{d y_1 \,d y_2  d^2 p_{T1} d^2 p_{T2}}\,\,\propto\,\,1
 \,\,+\,\,2 \sum_n v_{ n,n } \Lb p_{T1}, p_{T2}\Rb \,\cos\Lb n\,
\varphi\Rb
     \eeq
     where $ \varphi$ is the angle between    $\vec{p}_{T1}$ and
 $ \vec{p}_{T2} $.
     $v_n$ is determined  from  $v_{n,n}   \Lb p_{T1}, p_{T2}\Rb $
     \beq \label{vn}  
 1.~~    v_n\Lb p_T\Rb\,\,=\,\,\sqrt{v_{n,n}\Lb p_T, p_T\Rb}\,;\,
~~~~~~~~~~~~~2.~~~~  v_n\Lb p_T\Rb\,\,=\,\,\frac{v_{n, n}\Lb p_T,
 p^{\rm Ref}_T\Rb}{\sqrt{v_{n,n}\Lb p^{\rm Ref}_T, p^{\rm Ref}_T\Rb}}\,;
     \eeq
 \eq{vn}-1 and \eq{vn}-2  depict   two methods  of how the  values
 of $v_n$ have been extracted from the experimentally measured
  $v_{n,n} \Lb p_{T1}, p_{T2}\Rb$. $ p^{\rm Ref}_T$ denotes the
 momentum of the reference trigger.
     These two definitions are equivalent if  $v_{n, n}\Lb p_{T1},
 p_{T2}\Rb $ can be factorized as $v_{n, n}\Lb p_{T1}, p_{T2}\Rb\,=\,
     v_n\Lb p_{T1}\Rb\,v_n\Lb p_{T2}\Rb $.

     
\begin{boldmath}
     \section{Symmetry $\varphi \,\to\,\pi - \varphi$ ( $v_n = 0$ for odd  
$n$)
 for different multiplicities of  produced hadrons}     
 \end{boldmath}
 
 \begin{boldmath}     
\subsection{The Bose-Einstein correlation function   for deuteron-deuteron
 scattering with
 the correlation length  $L_c\,\,\propto\,\, R_D$}
 \end{boldmath}

\begin{figure}[ht]
 \includegraphics[width=10cm]{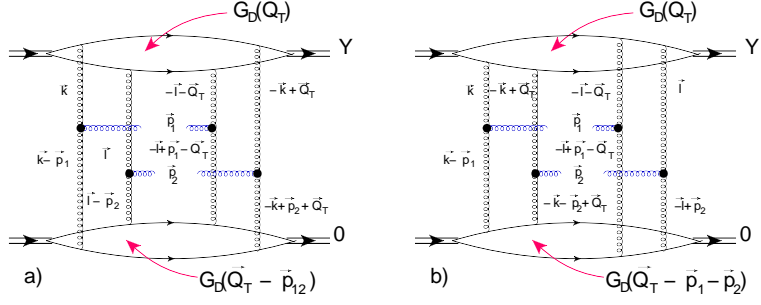}
   \protect\caption{ Deuteron-deuteron scattering in the Born approximation of
 perturbative QCD:
 \fig{ddba}-a describes the interference diagrams in the production of 
two
 identical gluons, in
 the process of multiparticle generation that gives rise to the correlation 
function
 $C\Lb L_c |\vec{p}_{12,T} = \vec{p}_{1,T} - \vec{p}_{2,T}|\Rb$; \fig{ddba}-b
  corresponds
 to  the central diffraction of  two  gluons with different colour charges in
 the colourless
 state. }
\label{ddba}
   \end{figure}

First, we consider the simplest diagram in the Born approximation
 of perturbative QCD, which
 we have discussed in Ref.\cite{GOLE} (see \fig{ddba}-a) ).
 This diagram describes the
 interference between two identical gluons in the process of multiparticle
 production, or
 in other words, in the processes of the production of two parton showers. 
 In this
 diagram
 $Q_T \propto 1/R_D$ and $ |\vec{Q}_T - \vec{p}_{12,T}| \propto 1/R_D$,  
 where $R_D $ 
denotes the
 deuteron radius, which is much larger than the size of the proton, $R_N$.
 Momenta $k_T$ , 
$l_T$,
 $p_{1,T}$ and $p_{2,T}$ in this diagram are of the order of
 $1/R_N \,\gg\,1/R_D$ and,
 therefore, we can neglect $Q_T$ as well as $p_{12,T}$, in the
 diagram.
Bearing this in mind,  we see that the correlation function
 $C\Lb L_c |\vec{p}_{12,T}|\Rb$ is
 equal to
\beq \label{BEDD1}
C\Lb L_c |\vec{p}_{12,T}|\Rb\,\,=\,\,\frac{1}{N^2_c - 1}\frac{\int d^2
 Q_T\, G_D\Lb Q_T\Rb\,G_D\Lb \vec{Q}_T - \vec{p}_{12,T}\Rb}{\int d^2\, Q_T G_D\Lb Q_T\Rb\,G_D\Lb Q_T\Rb}~~~~\mbox{with}~~~  G_D\Lb Q_T\Rb\,\,=\,\,\int d^2 r\,e^{i \vec{r} \cdot \vec{Q}_T}|\Psi_D\Lb r\Rb|^2\,
\eeq
 where $r$ denotes the distance between the proton and the neutron 
in the deuteron.

\eq{BEDD1} displays no symmetry with respect to $\varphi \,\to\,\pi -
 \varphi$. However,  we 
can 
add a different diagram of \fig{ddba}-b, which describes the central
 diffraction production 
of
 two different gluons in a colourless state \footnote{We are grateful
 to Alex Kovner and Michael Lublinsky who 
drew our attention to this diagram, and explained that in their
 approach \cite{KOLUCOR} this
 diagram restores the symmetry $\phi \,\to\,\pi - \phi$.}. 
 This diagram depends on
 $\vec{p}_{1,T} \,+\,\vec{p}_{2,T}$  and generates the correlation function
\beq \label{BEDD2}
\widetilde{C}\Lb L_c |\vec{p}_{1,T}\,+\,\vec{p}_{2,T}|\Rb\,
\,\propto\,\,\frac{1}{N^2_c - 1}\frac{
\int d^2 Q_T\, G_D\Lb Q_T\Rb\,G_D\Lb \vec{Q}_T - \vec{p}_{1,T}
 - \vec{p}_{2,T}\Rb}{\int d^2\, Q_T
 G_D\Lb Q_T\Rb\,G_D\Lb Q_T\Rb}
\eeq
since in this diagram $Q_T$ and $\vec{Q}_T - \vec{p}_{1,T} - 
\vec{p}_{2,T}$ are of the order
 of $1/R_D$, while $k_T$ , $l_T$, $p_{1,T}$ and $p_{2,T}$ in
 this diagram are of the order 
of
 $1/R_N \,\gg\,1/R_D$, therefore, we can neglect $Q_T$ as well
 as $\vec{p}_{1,T} +
 \vec{p}_{2,T}$ in the diagram or, in other words, we can put 
  $\vec{p}_{1,T} = -
 \vec{p}_{2,T}$.  After this substitution, both
 diagrams have the same expressions.

 Therefore, if  diagrams of \fig{ddba}-a and \fig{ddba}-b have the
 same weight, the sum will
 have the symmetry with respect to $\vec{p}_{2,T} \,\to\,- \vec{p}
_{2,T}$, restoring the
 symmetry with respect to $\varphi \,\to\,\pi - \varphi$. At first sight this
 is the case,
 since all integrations over $k_T$ and $l_T$ look the same. 
 However, in these two diagrams this is certainly not
 the case due to different integration with respect to
 $k_- $ and $l_-$ (or $k^+$ and $l^+$
 ). These integrations generates $1/4$
 suppression of the diagram of
 \fig{ddba}-b with respect to the diagram of \fig{ddba}-a. It is a well
 known fact, which for the first time, 
has
 been discussed  in the AGK paper of Ref.\cite{AGK}, as
 well as in the 
most
 reviews and books that are devoted to the high energy scattering 
 ( in particular those, 
  where one of us is an author \cite{GLR,GLR1,KOLEB}). For the completeness 
of
 presentation we add  appendix A in which we discuss this integration.

\begin{figure}[ht]
 \includegraphics[width=14cm] {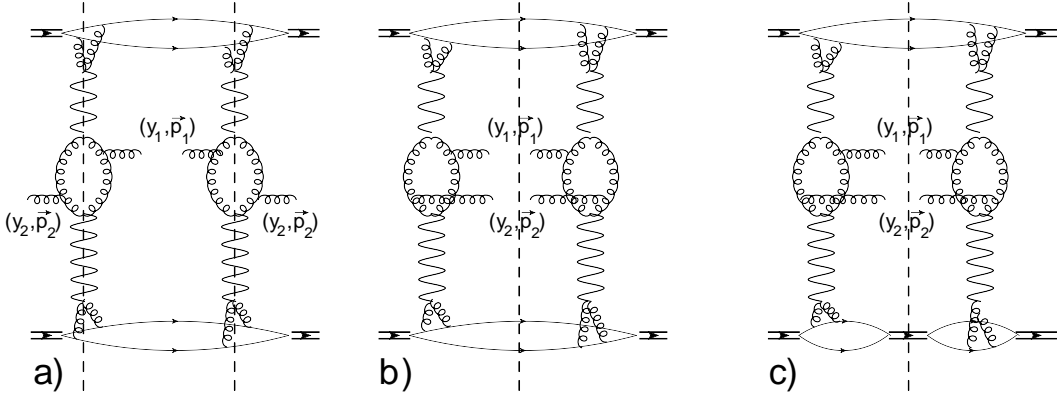}
   \protect\caption{ Mueller diagrams\cite{MUDI} for two parton shower
 production of  gluons: \fig{ddpom}-a describes the interference diagrams
 in the production of two identical gluons in the process of multiparticle
 production  that generates the correlation function $C\Lb L_c |\vec{p}_{12,T}
 \Rb$; \fig{ddpom}-b corresponds to central diffraction of  two  gluons with
 different colour charges in the colourless state; \fig{ddpom}-c describes the
 central diffractive production with a different final state, where one 
deuteron remains intact. 
The wavy line  stand for the BFKL Pomeron \cite{BFKL}. Helical 
lines correspond to gluons. The vertical dashed lines show the cuts.}
\label{ddpom}
   \end{figure}

 However, we found it  instructive to discuss the contribution
 of these two diagrams in 
the
 framework of the AGK cutting rules, which is the technique that we will use 
in considering
 the dependence of the correlation function  on multiplicity of
 produced particles. First, 
accounting for emission of the gluons with rapidities larger
 than $y_1$ and smaller than
 $y_2$, and considering $\bas |y_1 - y_2| \,\ll\,1$, we can describe
 the two partonic showers
 contribution in deuteron-deuteron scattering by the diagrams of
 \fig{ddpom}-a and \fig{ddpom}-b.

\begin{figure}[ht]
 \includegraphics[width=12cm]{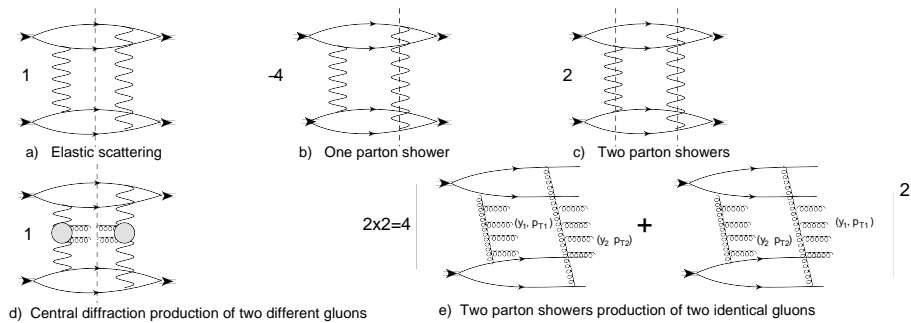}
   \protect\caption{ AGK  cutting rules for the exchange of two
BFKL Pomerons (\fig{agk}(a-c)) and the contributions of the central
  (\fig{agk}-d) and two parton showers production (\fig{agk}-e) of two gluons. }
\label{agk}
   \end{figure}

 The AGK cutting rules  describe the relative contributions of 
different processes 
that
 stem from two BFKL Pomeron \cite{BFKL,LI} exchange.
 \fig{agk}-a describes the elastic scattering, \fig{agk}-b  the one 
parton
 shower production,
 that is screened by the BFKL Pomeron exchange.
 \fig{agk}-c is
 the production of two parton showers.  The AGK cutting rules state that the 
cross
 sections of these three processes are related as $1 \,:\,-4\,:\,2$.
 The sum of 
these
 processes is equal to -1, leading to the negative contribution to the total 
cross
 section of two Pomeron exchange.
  These rules  have a rather general origin based on the unitarity
 constraints  and  physical properties of the Pomerons.
 Indeed, the unitarity constraint has the following form
 \beq \label{UNIT}
 2 \,\mbox{Im} A_{el}\Lb s, b; i \Rb\,\,=\,\,\underbrace{| A_{el}\Lb s, b; i
 \Rb|^2 }_{\rm elastic \,\,cross\,\,section}\,\,+\,\,
\underbrace{G\Lb s, b,i\Rb}_{\rm
 contribution \,of\, inelastic\,processes}
 \eeq
  where $W = \sqrt{s}$   denotes the energy of the collision, $b$ is 
 impact parameter,
 and $i$  the set of other quantum numbers that diagonalize the
 interaction matrix.

 For the BFKL Pomeron, the elastic cross section is much smaller than the 
exchange of
 a single Pomeron, and \eq{UNIT} takes the form
 \beq \label{UNIT1}
  2 \,\mbox{Im} P^{\rm BFKL}\Lb  s, b,i\Rb\,\,=\,\,
\underbrace{G^{\rm BFKL}\Lb s, b,i\Rb}_{\rm cut\,\,Pomeron}
  \eeq
 
 Using \eq{UNIT1} one can see that
 \bea \label{UNIT2}
 &&\sigma_{el} \,\propto\, | P^{\rm BFKL}\Lb  s, b,i\Rb |^2;~~~\sigma_{
\rm one\,
 parton\,shower}\,\propto\,- \,2 \,P^{\rm BFKL}\,G^{\rm BFKL}\Lb s,
 b,i\Rb;\nn\\
&&~~~\sigma_{\rm two \, parton\,showers} \,\propto\,- \,\h \,G^{\rm
 BFKL}\Lb s,
 b,i\Rb\,G^{\rm BFKL}\Lb s, b,i\Rb; 
 \eea
 where $\h$ in the last term stem from the fact that the two cut Pomerons 
are identical.
 Using
 \eq{UNIT1} one reproduces the AGK cutting rules of \fig{agk}-a - \fig{agk} -c.

 The central diffraction  production of two gluons 
is shown in
 the diagram of 
\fig{agk}-a (elastic scattering), while the interference diagram, that generates
 the Bose-Einstein correlations, originates from \fig{agk}-c with the extra factor
  2, which reflects the fact that the gluon with rapidity, say, $y_1$ can be 
produced
 from two different parton cascades (see \fig{agk}-e). The
 processes of central diffractive production are  suppressed by a factor of 4 
 compared to
  the Bose-Einstein correlations.
 
 To complete the discussion of the possible restoration of $\varphi 
\,\to\,\pi -
 \varphi$ symmetry, due the processes of the central diffraction, we note that
 in these processes there can be a final state in which one or two deuterons 
remain
 intact ( see for example \fig{ddpom}-c) which leads to  different correlation 
functions. For example for \fig{ddpom}-c the correlation function  has 
the  form
 \beq \label{BEDD3}
C_{\rm \fig{ddpom}-c}\Lb L_c |\vec{p}_{1,T}\,+\,\vec{p}_{2,T}|\Rb\,\,\propto\,
\,\frac{1}{N^2_c - 1}\frac{\int d^2 Q_T\, G_D\Lb Q_T\Rb\,G^2_D\Lb \vec{Q}_T - 
\vec{p}_{1,T} - \vec{p}_{2,T}\Rb}{\int d^2\, Q_T G_D\Lb Q_T\Rb\,G_D\Lb Q_T\Rb}
\eeq 
 which  differs from \eq{BEDD2}.
 
 A comment regarding the status of the AGK cutting rules in QCD. For 
 deuteron-deuteron scattering, the cutting rules  shown in 
\fig{agk}-a -  \fig{agk}-c ,
 have been proved on  general grounds\cite{AGKDD}, using unitarity
 and the wave nature of the colliding particles.  In the framework 
of perturbative 
QCD these cutting rules were proven in Refs.\cite{GLR,AGK0}. For the
 inclusive cross
 sections, the AGK cutting rules were discussed and proven in 
Refs.\cite{AGK1,AGK2,AGK3,AGK4,AGK5,AGK6,AGK7}. However, in Ref.\cite{AGK8} it
 is shown that the AGK cutting rules are violated for  double inclusive
 production. This violation is intimately related to the enhanced diagrams
 \cite{AGK7,AGK8}, and reflects the fact that different cuts of the triple BFKL 
Pomeron
 vertex  lead to different contributions. Recall, that we do not
 consider such diagrams.

\begin{figure}[ht]
 \includegraphics[width=14cm]{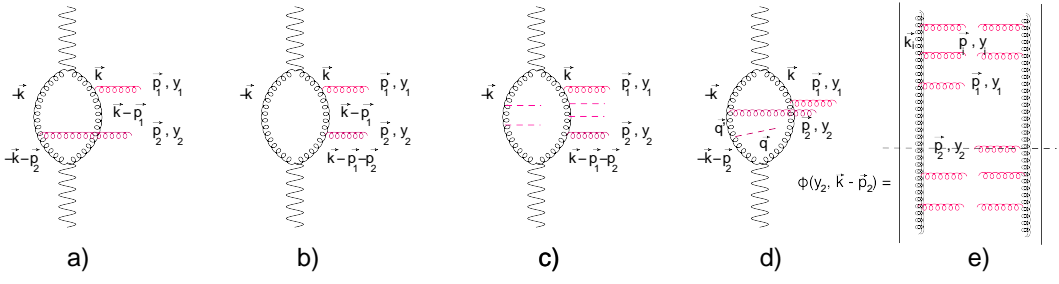}
   \protect\caption{ Vertex for emission of two gluons by the BFKL Pomeron.
 \fig{vG}-c shows the emission of soft gluons  whose suppression 
leads to the
 Sudakov form factor. \fig{vG} shows the emission of the gluon in the
 DLA approximation of perturbative QCD, which leads to the Sudakov
 form factor in the vertex of two gluons emission .}
\label{vG}
   \end{figure}

  Therefore, the contribution of the central diffraction process is
 suppressed by a factor of  four, due to the longitudinal momenta 
integration. However,
 we need to compare the values of the vertices for gluon inclusive
 production (see \fig{agk}-d )and the vertex for two gluon production
 from the BFKL Pomeron.  From  \fig{vG} we can see that this
 vertex, is two times larger than the vertex for gluon inclusive 
production.
 Indeed, the contribution of \fig{vG}-a is the same as for inclusive
 production, but we have to add \fig{vG}-b. In appendix B we show that
 these two diagrams (\fig{vG}-a and \fig{vG}-b) are the same. Adding these
 diagrams we note that for deuteron-deuteron scattering we expect, the
 symmetry $\varphi \to \pi - \varphi$ in the measurements with no selection
 on multiplicity. This observation supports  the claim of
 Refs.\cite{KOWE,KOLUCOR}. 
 
 In this paper as well as in Refs.\cite{KOWE,KOLUCOR,GOLE} we discuss the 
case $\bas | y_1 - y_2| \leq 1$. Let us consider this restriction  more 
carefully. 
   We start with writing the expression for  the two 
diagrams of
 \fig{agk}-d.  The inclusive cross section for production of the gluon
 with rapidity $y_1$ and transverse momentum $p_{1,T}$ due to the 
exchange  of one BFKL Pomeron, has the following form

  \beq \label{VG1}
 \frac{d \sigma}{d y_1 \, d^2 p_{T1} }\,\,\propto\,\,\frac{\bas}{p^2_{1,T}}\int d^2 k_T  \, \phi^{\rm BFKL}\Lb Y - y_1,k_T\Rb\frac{\Gamma_\mu\Lb k_T, p_{1,T}\Rb\, \Gamma_{\mu}\Lb  k_T, p_{1,T}\Rb}{k^2_T\, \Lb \vec{k}_T - \vec{p}_{1,T}\Rb^2}\,\phi^{\rm BFKL}\Lb  y_1,k_T\Rb  
 \eeq

 The interference diagram in which the parton shower  with a gluon
 with $y_1$ and $p_{1,T}$ in the amplitude, is squared with the parton
 shower in which a gluon with $y_2$ and $p_{2,T}$ is produced, takes the 
form
   \beq \label{VG2}
 \frac{d \sigma}{d y_1 \, d^2 p_{T1} }\,\,\propto\,\,\frac{\bas}{p^2_{1,T}}
\int d^2 k_T  \, \phi^{\rm BFKL}\Lb Y - y_1,k_T\Rb
 \frac{\Gamma_\mu\Lb k_T, p_{1,T}\Rb\, \Gamma_{\nu}\Lb 
 k_T, p_{2,T}\Rb}{k^2_T\, \Lb \vec{k}_T - \vec{p}_{2,T}\Rb^2}\,
\phi^{\rm BFKL}\Lb  y_2,k_T\Rb  
 \eeq
  In \eq{VG1} and \eq{VG2} we neglected $p_{12,T} \propto 1/R_D$ as we
 have explained above.
  
   In \eq{VG1} and \eq{VG2} $\phi$ is the solution of the BFKL equation
   \beq \label{BFKL}
\frac{\partial \phi^{\rm BFKL}\Lb y, \vec{k}_T \Rb}{\partial y}\,=\,\bas
 \int \frac{d^2 k'_T}{ \pi}\,\frac{1}{\Lb \vec{k}_T - \vec{k'}_{T}\Rb^2}
\,\phi^{\rm BFKL}\Lb y, \vec{k'}_T\Rb\,\,-\,\,2  \omega_G\Lb  \vec{k}_T\Rb
\,G\Lb y, \vec{k}_T\Rb\,;\eeq
where
\beq \label{OMEGA}
\omega_G\Lb\vec{k}_T\Rb = \h \bas k^2_T \int \frac{d^2 k'_T}{2 \pi}
 \frac{1}{k'^2_T\,\Lb \vec{k}_T - \vec{k'}_{T}\Rb^2 }\,=\,  \bas k^2_T 
\int \frac{d^2 k'_T}{2 \pi} \frac{1}{\Lb k'^2_T\,+\,\Lb \vec{k}_T - \vec{k'}_{T}\Rb^2\Rb\,\Lb \vec{k}_T - \vec{k'}_{T}\Rb^2}
\eeq

 Comparing \eq{VG1} and \eq{VG2} one can see that  to neglect the 
difference
 between $y_2$ and $y_1$ in $\phi^{\rm BFKL}\Lb  y_2,k_T\Rb$
 we need to assume
 that $ 2.8\, \bas\, |y_1 - y_2| \ll\,1$
  ( $2.8\,
 \bas$ is the intercept of the BFKL Pomeron). However, the actual
 restriction turns out to be even more severe. Indeed, in all
 interference diagrams as well as in double gluon production between
 rapidities $y_1$ and $y_2$, we have the exchange  in the $t$-channel 
of two gluons in the octet state.
 This  means that we have
 the additional emission of gluons with rapidities between $y_1$ and
 $y_2$ (see \fig{vG}-c). This emission leads to the extra Sudakov form
 factor\cite{DURG} in \eq{VG2} which takes the form:
  \beq \label{VG3}
 \frac{d \sigma}{d y_1 \, d^2 p_{T1} }\,\,\propto\,\,\frac{\bas}{p^2_{1,T}}
\int d^2 k_T  \,e^{-\,S\Lb \delta y , k_T, p_{1,T}\Rb}\, \phi^{\rm BFKL}\Lb
 Y - y_1,k_T\Rb
 \frac{\Gamma_\mu\Lb k_T, p_{1,T}\Rb\, \Gamma_{\nu}\Lb
  k_T, p_{2,T}\Rb}{k^2_T\, \Lb \vec{k}_T - \vec{p}_{2,T}\Rb^2}
\,\phi^{\rm BFKL}\Lb  y_2,k_T\Rb  
 \eeq
 where $\delta Y = |y_1 - y_2|$. We recall the structure of the
 one parton shower that is described by the BFKL Pomeron in
 \fig{vG}-e \cite{BFKL}, the one parton shower is given by 
 \beq \label{VG4} 
 \prod^n_{i=1}\Gamma_{\mu}\Lb k_{i,T}, p_{i,T}\Rb 
\,\frac{e^{\omega_G\Lb k_{i,T}\Rb\,\Lb y_i - y_{i - 1}\Rb}}{k^2_{i,T}}
 \eeq
 which being squared, leads to the parton density 
$\phi\Lb y, k_{1,T}\Rb$. In simple words the BFKL
 cascade is the ladder diagram with specific vertices
 of gluon production, and with the exchange of the reggeized
 gluons with trajectories which are given by \eq{OMEGA}.
 Absorbing the terms in $\phi(y, k_T )$  for \eq{VG3} we see that
 \beq \label{S}
 S\Lb \delta y, k_T,p_{1,T} \Rb\,\,=\,\Lb \omega\Lb \vec{k}_T -
 \vec{p}_{1,T}\Rb + \omega \Lb \vec{k}_T\Rb\Rb \delta y\,=
\,\frac{\bas}{2}\Lb \ln\Lb \Lb \vec{k}_T - \vec{p}_{1,T}\Rb^2/\mu^2\Rb\,+
\,\ln\Lb k^2_T/\mu^2\Rb\Rb \,\delta y
 \eeq
 and it has a typical Sudakov form factor structure. $\mu$ is the
 typical dimensional parameter which in the DGLAP evolution, is of
 the order of  the soft scale in the hadron, and in CGC it is a saturation
 scale $Q_s\Lb y_1\approx   y_2\Rb $.
 
  For the diagrams of \fig{vG}-a and \fig{vG}-b we  need to introduce 
the same suppressions.  These Sudakov suppressions  result 
from the fact that
 in  the  approximation  for $\bas \delta y\,\ll\,1$  we take into 
account
 only simple diagrams with two gluons, and  without extra  gluon 
emissions;
 and they  stipulate the size we need to take  for $\delta y$. 
However, the  two
 gluon production has  an additional suppression of  the   
Sudakov type , which applies even at $y_1 =  y_2$,
 where $S$ of \eq{S} is equal to zero: the emission of gluons that
 are shown in \fig{vG}-d,  has been discussed in detail in
 Ref.\cite{DDT,DURG}.
 
   This emission leads to the value of $S$ in the  double log 
approximation of perturbative QCD that has the form:
   \beq \label{DLAS}
   S\Lb p_{1,T}, k_T\Rb\,\,=\,\,\frac{\bas}{\pi} \int^{M/2}_{k_T}
\frac{d^2 q_T}{q^2_T} \int^{M/2}_{q_T} \frac{d q_0}{q_0}\,\,=\,
\,\frac{\bas}{4}\,\ln^2\Lb \frac{M^2}{4\,k^2_T}\Rb
   \eeq
    where $M$ denotes the mass of the produced dijet,  which is given 
by $M^2
 \,=\,2 p^2_T\Lb 1 + \cosh\Lb y_1 - y_2\Rb\Rb$ considering $\vec{p}_{1,T} = -
 \vec{p}_{2,T} = \vec{p}_T$.
     The limits in integration over $q_0$ can easily be understood 
in the rest frame of the two gluon jets.   In this frame 
 the minimal $q_0 = q_T$. The lower limit in $q_T$ integration
 stems from the fact, that at distances less than $1/q_T$, the emission 
with 
two $t$-channel gluons have a distructive interference canceling the 
emission,
 since the total colour charge is zero.  For $ q_T \,\geq\,q_T$ the
 emission of  gluons comes from the   $t$-channel gluon, which carries
 color, and leads to the color coefficient in \eq{DLAS}.

 Finally, the contribution of the
 diagram of \fig{vG}-b has   the following for $y_1 = y_2$:
   \bea \label{VG5}
&& \frac{d \sigma}{d y_1 \, d^2 p_{T1} }\Lb \fig{vG}-b\Rb\,\,\propto\,\\
&&\,\frac{\bas}{p^2_{1,T}}\int d^2 k_T  \,e^{-\,S\Lb M, k_T\Rb}\,
 \phi^{\rm BFKL}\Lb Y - y_1,k_T\Rb
 \frac{\Gamma_\mu\Lb\vec{k}_T, \vec{p}_{1,T}\Rb\, \Gamma_{\nu}\Lb   \vec{k}_T
  - \vec{p}_{1,T}, \vec{p}_{2,T}\Rb}{k^2_T\, \Lb \vec{k}_T - 
\vec{p}_{1,T}\Rb^2}\,\phi^{\rm BFKL}\Lb  y_1, k_T \Rb   \nn
 \eea

\begin{figure}[ht]
 \includegraphics[width=8cm]{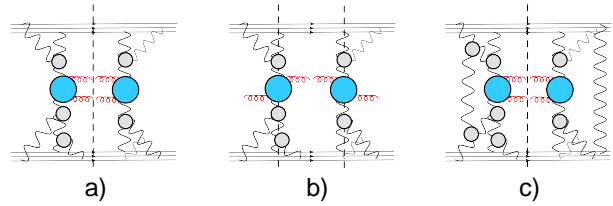}
   \protect\caption{ The double inclusive production for dense-dense parton
 system scattering: the central diffraction production (see \fig{froi}-a) and
 the Bose-Einstein correlation of the identical gluons ( \fig{froi}-b). The wavy
 lines denote the BFKL Pomerons. \fig{froi}-c shows  the diagrams that
 do not contribute for the inclusive production of two gluons. The 
 green blobs show the Mueller vertices for 
two gluon production, while the circles stand for the triple Pomeron vertices. 
 The produced  gluons are denoted by red helical lines.}
\label{froi}
   \end{figure}

  The integration over  $k_T$ of the parton densities is concentrated
 in the vicinity of the saturation scale, since in coordinate space
 $\phi \propto \nabla^2 N\Lb r, y\Rb $\cite{KOTUINC}, deep in the 
saturation region  it tends to zero.
  Of course,  we consider 
not only one BFKL Pomeron, but a more complicated structure of the single
 parton cascade (see \fig{froi}).  Therefore, substituting $Q_s$ instead
 of $k_T$ in the Sudakov form factor, we  find that \eq{VG5}    takes the 
form:
 
    \bea \label{VG6}
&& \frac{d \sigma}{d y_1 \, d^2 p_{T1} }\Lb \fig{vG}-b\Rb\,\,\propto\,\,\frac{\bas}{p^2_{1,T}}\exp\Lb - \frac{\bas}{4} \ln^2\Lb \frac{p^2_T\Lb 1 + \cosh\Lb y_1 - y_2\Rb\Rb}{2\,Q^2_s\Lb y_1\approx y_2\Rb}\Rb\Rb\\
&&\int d^2 k_T  \,\, \phi^{\rm BFKL}\Lb Y - y_1,k_T\Rb
 \frac{\Gamma_\mu\Lb\vec{k}_T, \vec{p}_{1,T}\Rb\, \Gamma_{\nu}\Lb   \vec{k}_T  - \vec{p}_{1,T}, \vec{p}_{2,T}\Rb}{k^2_T\, \Lb \vec{k}_T - \vec{p}_{1,T}\Rb^2}\,\phi^{\rm BFKL}\Lb  y_1, k_T \Rb   \nn
 \eea   
 
 However, for discussing the current experimental data, especially
 for hadron-hadron interactions,  for the parton densities,  we can use
 the experimental data for DIS structure function which  is  well
 described \cite{HERA}, by the DGLAP evolution equations \cite{DGLAP}.
 In this case,  we need to put the value of $Q_0 = Q_s\Lb y_1 = Y_0 \approx
 3\Rb$ from the Colour Glass Condensate (CGC) motivated fit of HERA
 data\cite{AAMQS,IMMST}. This value turns out to be in the range
 $ Q_0 = 0.2 - 0.5 \,GeV$\cite{AAMQS,IMMST}.
 
 Finally, we obtain  the resulting correlation function is the
 sum of \eq{BEDD2} and \eq{BEDD3}:
 \beq \label{CORSUM}
 C\Lb \varphi\Rb\,=\,C\Big( \eq{BEDD2};{\rm  L_c \,2\, p_T \,\sin(\varphi)
 }\Big) \,+\,e^{- \frac{\as}{2}\ln^2\Lb \frac{p^2_T\Lb 1 + \cosh\Lb y_1 - y_2\Rb\Rb}{2\,Q^2_s}\Rb}\,C\Big(
 \eq{BEDD3};{\rm L_c\, 2\, p_T \,\cos(\varphi) }\Big) 
 \eeq
 where we assume that $|\vec{p}_{1,T}| = |\vec{p}_{2,T}| = p_T$.
  
  The general expectation from \eq{CORSUM}  indicates  that $v_n$ with 
odd $n$ will peak
 at $p_T \approx 4 Q_0$, where the second term will be  approximately three
 times  smaller that the first one.  The experimental data for $v_n$ 
in 
proton-proton collisions\cite{ATLASVNPP} show that
 $v_n$ reaches a maximum at $p_T \approx 3 \,GeV$,  and this value 
 is independent of
 the  energy.  Such a behaviour 
 qualitatively supports \eq{CORSUM} with $Q_0 \approx 0.6\,GeV$.
  
     Concluding this section we would like to summarize our results:
 (i) we showed that at small transverse  momenta the processes of
 exclusive(diffractive)  in the central rapidity region (CED)  of
 two gluons,  are equal to the interference contributions of two parton
 showers, confirming the results of Refs.\cite{KOWE,KOLUCOR}, this fact
 leads to $v_n = 0$ for odd $n$, in the total inclusive measurements, 
without
 any selection on multiplicity of produced hadrons;
 (ii) we found the mechanism of suppression of CED of two gluon
 jets for large transverse momenta due to Sudakov form factor,
 which leads to the correlation function of \eq{BEDD2}, and to
 $v_n \neq 0$ for odd $n$, in the experiments without selection
 on multiplicities; and (iii) only the correlation function of
 \eq{BEDD1} can be measured in the processes of multiparticle
 generation with the multiplicities $N\geq \bar{n}$, 
 where $\bar{n}$ is the average multiplicity in the collisions. The process
 of the central diffraction which generates the correlation function of
 \eq{BEDD2} corresponds to the event with low multiplicity 
$N\,<\,\bar{n}$).
 The last item is the best motivation for study of the identical
 particle correlations  $v_n$  with even  $n$, and  with different 
multiplicities,
 which we will consider below.

\begin{figure}[ht]
 \includegraphics[width=10cm]{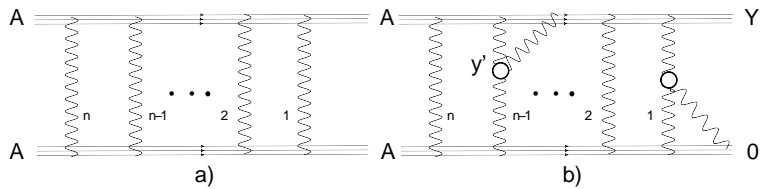}
   \protect\caption{  Nucleus-nucleus scattering in the Glauber\cite{ GLA}
 approach (\fig{gla}-a), and  the first corrections to this approach due to
  triple BFKL Pomeron interactions (\fig{gla}-b). The wavy lines denote
 the BFKL Pomerons. The blobs show the triple Pomeron vertices. }
\label{gla}
   \end{figure}

  \begin{boldmath}     
\subsection{Bose-Einstein correlation function   for heavy ions
  scattering with
 the correlation length  $L_c\,\,\propto\,\, R_A$}
 
 ~
 
 \subsubsection{Inclusive measurements}
 \end{boldmath}
Concluding the previous subsection, we claim that for deuteron-deuteron
 scattering, we  see how the processes of the central diffraction,
in the measurements
 that sum processes with all possible multiplicities of produced 
particles,
 can lead to the symmetry $\varphi\,\,\to\,\pi - \varphi$ for $p_T \leq 
Q_s$.
   In
 this section we would like to examine, if such symmetry could be possible  
for
 ion-ion interactions,  which  can be described by the Glauber \cite{GLA} 
formula
 (see \fig{gla}-a):
\beq \label{AAGLA}
A_{\rm AA}\Lb s, b\Rb\,\,=\,\,i \Lb 1 - \exp\Lb - \Omega\Lb s, b\Rb\Rb\Rb
~~~\mbox{with}~~~\Omega\Lb s, b \Rb\,\,=\,\,g^2_A \,P^{\rm BFKL}\Lb s,b\Rb\,
T_{\rm AA}\Lb b\Rb
\eeq
where $T_{\rm AA}\Lb b \Rb$ is the optical width  and given by 
\beq \label{AAGLA1}
T_{\rm AA}\Lb b \Rb\,\,=\,\,\int d^2 b' \,S_A\Lb \vec{b} - \vec{b}'\Rb\,S_A\Lb
 \vec{b}'\Rb~~~~~~\mbox{with}~~~ S_A\Lb b \Rb\,\,=\,\,\int ^{+\infty}_{-\infty}
 d z \rho\Lb z, \vec{b}\Rb~~~~~~\int d^2 b \,S_A\Lb b \Rb\,\,=\,\,A
\eeq
where $\rho\Lb z, \vec{b}\Rb$ denotes the nucleon density in the nucleus,
 and $z$
  the longitudinal coordinate of the nucleon. In \eq{AAGLA} $g_N$ denotes the 
impact factor that describes the interaction of the BFKL Pomeron  (whose
 Green function  is $P^{\rm BFKL} $), with the nucleon.

We wish to stress that \eq{AAGLA} in the framework of perturbative QCD 
(pQCD) has
 a region of applicability. Indeed, the contribution of one BFKL Pomeron
in pQCD, in
 \eq{AAGLA}, is proportional  to $g^2_N P^{\rm BFKL}\,T_{\rm AA}\Lb b\Rb 
\,\propto\,\,\bas^2\,A^{4/3} \exp\Lb \Delta_{\rm BFKL} Y\Rb $ where $\Delta_{\rm
 B FKL} \,\propto\,\bas$,  where $\Delta_{\rm BFKL}$ denotes the BFKL Pomeron 
intercept. The first `fan' diagrams
  lead to  corrections to the Glauber formula, these  are shown in
 \fig{gla}-b , and are of the order 
\beq \label{AAGLA2}
g^2_N P^{\rm BFKL}\Lb Y \Rb\,T_{\rm AA}\Lb b\Rb\,\int^Y_0 d y' G_{3\pom}
 \,g_N\,P^{\rm BFKL}\Lb y' \Rb\,S_A\Lb b \Rb\,\,\propto\,\,\bas^4\,A\,\Lb
 P^{\rm BFKL}\Lb Y,b\Rb\Rb^2
\eeq
Comparing \eq{AAGLA2} with the exchange of two BFKL Pomerons, we see that 
 the contribution of the fan diagrams will be smaller that 1 for
 $\Delta_{\rm BFKL} Y\,\,\ll\, \,\h \ln\Lb1/(\bas^4\,A)\Rb$, while
 the contribution of the BFKL Pomeron in Glauber formula will be larger
 than 1. In other words, for $Y \,\leq\, \Lb 1/(2\,\Delta_{\rm BFKL} )\Rb
\,\ln\Lb1/(\bas^4\,A)\Rb$ we can describe the ion-ion collisions using the
 Glauber formula of \eq{AAGLA}.

In this formula the contributions of $n$-BFKL Pomeron exchanges  to the 
total
 cross section is equal to
\beq \label{AAGLA3}
 \sigma^{(n)}_{\rm tot}\,=\,\frac{2 \,(-1)^{n - 1}}{n!} \Omega^n\Lb s,b\Rb.
 \eeq  
 Accordingly to the AGK cutting rules,  the relative weight of the process with
 $m$ cut Pomerons, ($n - m+1$ of them are not cut) is equal
\beq \label{AGK}
\frac{\sigma^{(m)}_n}{\sigma^{(n)}_{\rm tot}}\,\,=\,\,\Lb - 1 \Rb^{n - m}
 \,\frac{n!}{m!\,(n - m)!} \,2^{n-1}\,~~\mbox{for}~~m\,\geq\,1;~~~~~~~~
\frac{\sigma^{(0)}_{n}}{\sigma^{(n)}_{\rm tot}}\,\,=\,\,\Lb - 1
 \Rb^n\Lb 2^{n-1}\, - 1\,\Rb  ;
\eeq

\begin{figure}[ht]
 \includegraphics[width=10cm]{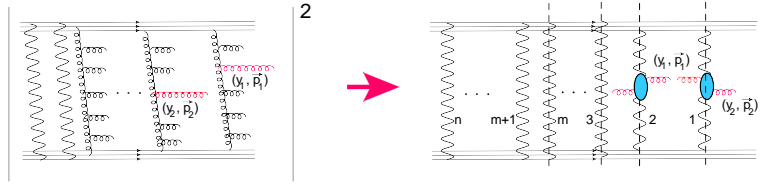}
   \protect\caption{ The contribution of different processes  of production
 of the number of parton showers (more than 2), to the Bose-Einstein
 correlation.
    The wavy lines denote the BFKL Pomerons. The blobs show the Mueller vertices
 for two gluon production .  The produced identical gluons are denoted in red
 helical lines.}
\label{2shbe}
   \end{figure}


To find the contribution of all possible processes of different multiplicities
 related to the production of $m$-parton showers, we need to calculate the
 following sum (see \fig{2shbe})
\bea \label{AAGLA4}
&& \frac{d^2 \sigma}{d y_1 \,d y_2  d^2 p_{T1} d^2 p_{T2}}\,\,=\\
&& C_A\Lb L_c |p_{12,T}|\Rb\,\frac{d \sigma^{\rm BFKL}}{d y_1
 \,d^2 p_{T1}}\,\frac{d \sigma^{\rm BFKL}}{d y_2 \,d^2 p_{T2}}
\,\sum^\infty_{m=2}\sum^n_{m=2}\,m (m-1)\,\frac{\sigma^{(m)}_n
 }{\Omega^2\Lb s,b\Rb} \,\,  =\,\,4 C\Lb L_c |p_{12,T}|\Rb\,\frac{d
 \sigma^{\rm BFKL}}{d y_1 \,d^2 p_{T1}}\,\frac{d \sigma^{\rm BFKL}}{d y_2
 \,d^2 p_{T2}}\nn
 \eea
In \eq{AAGLA4} we use \eq{AGK}, \eq{AAGLA3}, the  function $C_A$ is 
determined
 by an equation which is similar to  \eq{BEDD1}. Neglecting all 
correlations
 inside the nucleus, its wave function can be written as $\Psi_A\Lb \{r_i\}
\Rb\,\,=\,\,\prod^{A}_{i=1} \Psi_i\Lb r_i\Rb$ where $\Psi\Lb r_i\Rb$ 
denotes the 
wave function of $i$-th nucleon. In this approach
\beq \label{CA}
C_A\Lb L_c |p_{12,T}|\Rb\,\,=\,\,\frac{1}{N^2_c - 1} \frac{ \int d^2 Q_T\,G^2_A\Lb
 Q_T\Rb\,G^2_A\Lb \vec{Q}_T - \vec{p}_{12,T}\Rb}{ \int d^2 Q_T\,G^4_A\Lb Q_T\Rb}~~
~~~\mbox{with}~~~~G_A\Lb Q_T\Rb\,\,=\,\,\int d^2 b\, e^{i \vec{b} \cdot \vec{Q}_T}
\,S_A\Lb b \Rb
\eeq
where $S_A\Lb b\Rb$ denotes the number of the  nucleons at fixed impact parameter 
$b$.

\eq{CA} can be re-written in the impact parameter representation using \eq{AAGLA1}:
 viz.
\beq \label{CAB}
C_A\Lb L_c |p_{12,T}|\Rb\,\,=\,\,\frac{1}{N^2_c - 1}\frac{\int d^2 \tilde{b} \,e^{
 i \vec{\tilde{b}}\cdot \vec{p}_{12,T}} \,T^2_A\Lb \tilde{b}\Rb}{\int d^2 \tilde{b}
 \, \,T^2_A\Lb \tilde{b}\Rb}~~~~~~~~\mbox{where}~~~~~ T_A\Lb b \Rb\,=\,\int
 d^2 b'\,,S_A\Lb b' \Rb\,S_A\Lb \vec{b} - \vec{b}' \Rb\eeq

The production of gluons by the BFKL Pomerons given by the Mueller diagrams in
 \fig{2shbe},  generally has a more complicated form than we used in \eq{AAGLA4}
 ( see Eq.(38) of Ref.\cite{GOLE}),  and cannot be reduced to the production of
 single inclusive cross sections. However, in the case of deuteron
 scattering,
 we can consider $\vec{p}_{1,T} = \vec{p}_{2,T}$, since the difference $p_{12,T}
 \sim 1/R_D \,\ll\,1/R_N $ or $ \ll\,Q_s$, where $1/R_N$ and $Q_s$ are typical
 momenta in the BFKL Pomeron. Bearing this in mind, we can replace the 
contribution
 of the Mueller diagram by the single inclusive production of the gluon, 
by the BFKL Pomeron.

\begin{figure}[ht]
 \includegraphics[width=10cm]{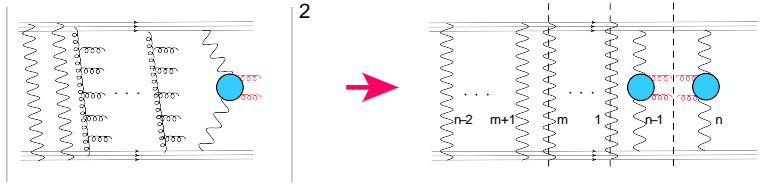}
   \protect\caption{ The contribution of different processes  of production of 
the number of parton showers to the central diffraction production.
    The wavy lines denote the BFKL Pomerons. The blobs show the Mueller vertices
 for two gluons production .  The produced  gluons are denoted  by  
red helical 
lines.}
\label{2shced}
   \end{figure}

 The contribution to the central diffraction productions is shown in
 \fig{2shced}, and takes the following form
 \bea \label{AAGLA5}
&&  \frac{d^2 \sigma}{d y_1 \,d y_2  d^2 p_{T1} d^2 p_{T2}}\,\,=\\
&&\,\, C_A\Lb L_c |\vec{p}_{1,T} + \vec{p}_{2,T}|\Rb\,\frac{d \sigma^{\rm
 BFKL}}{d y_1 \,d^2 p_{T1}}\,\frac{d \sigma^{\rm BFKL}}{d y_2 \,d^2 p_{T2}}
\,\Bigg( \sum^{n-2}_2 \frac{n!}{2! (n-2)!}\frac{ \sigma^{(0)}_n}{\Omega^2\Lb
 s,b\Rb}\,\,+\,\,\sum^\infty_{n=1}\sum^{n-2}_{m=1}\,\frac{n!}{2! (n - m - 2)!}
\,\frac{\sigma^{(m)}_n }{\Omega^2\Lb s,b\Rb} \Bigg)\,\, \nn\\
&& =\,\,C_A\Lb L_c |\vec{p}_{1,T} + \vec{p}_{2,T}|\Rb\,\frac{d \sigma^{\rm BFKL}}
{d y_1 \,d^2 p_{T1}}\,\frac{d \sigma^{\rm BFKL}}{d y_2 \,d^2 p_{T2}}\,\xrightarrow
{\Omega \gg 1} \,2\,C_A\Lb L_c |\vec{p}_{1,T} + \vec{p}_{2,T}|\Rb\,\frac{d \sigma^
{\rm BFKL}}{d y_1 \,d^2 p_{T1}}\,\frac{d \sigma^{\rm BFKL}}{d y_2 \,d^2 p_{T2}}\nn
 \eea

 In \eq{AAGLA5} we use \eq{AGK}, \eq{AAGLA3} and the function $C_A\Lb L_c 
|\vec{p}_{1,T}
 + \vec{p}_{2,T}|\Rb$ from \eq{CAB}, as well as $\vec{p}_{1,T} \,=\,- 
\vec{p}_{1,2} $
 for deuteron-deuteron scattering. Actually , these  estimates are correct only 
in the 
region of large $\Omega$. The general expression for the correlation function
 has the following form
 \bea \label{CAGB}
&&\hspace{-0.7cm}  C_A\Lb L_c |\vec{p}_{1,T} + \vec{p}_{2,T}|\Rb \,=\\
&&\hspace{-0.9cm}\,\frac{1}{N^2_c - 1}\frac{\int d^2 \tilde{b}\, e^{ i
 \vec{\tilde{b}}\cdot (\vec{p}_{1,T} + \vec{p}_{2,T})}\,\int  d^2 B \int
 d^2 b\,S_A\Lb \vec{B}  + \h \vec{\tilde{b}}\Rb S_A\Lb \vec{B}  - \h
 \vec{\tilde{b}}\Rb  S_A\Lb \vec{b}  + \h \vec{\tilde{b}}\Rb S_A\Lb \vec{b}
  - \h \vec{\tilde{b}}\Rb  \,\Bigg( 2 - \exp\Lb - \Omega \Lb \vec{b} + \vec{B}\Rb\Rb
\Bigg)}{\int d^2 \tilde{b} \int  d^2 B \int d^2 b\,S_A\Lb \vec{B}  + \h 
\vec{\tilde{b}}\Rb S_A\Lb \vec{B}  - \h \vec{\tilde{b}}\Rb  S_A\Lb \vec{b} 
 + \h \vec{\tilde{b}}\Rb S_A\Lb \vec{b}  - \h \vec{\tilde{b}}\Rb  \,\Bigg( 2
 - \exp\Lb - \Omega \Lb \vec{b} + \vec{B}\Rb\Rb\Bigg)  }\nn
  \eea
  We also make use of the fact that the Mueller vertex for production of two 
gluons by
 the BFKL Pomeron (see \fig{2shced}), is equal to the Mueller vertex 
 for inclusive production of a single gluon (see \fig{2shbe}).
 
 Comparing \eq{AAGLA4} and \eq{AAGLA5} we see that the contribution of
 the central diffraction production, is  twice  as large (at small 
$p_T$) than the contribution
 of the Bose-Einstein correlations. Therefore, the dominant contribution
 comes from \eq{AAGLA5} leading to the negative values of  $v_{n, n}$
 for  odd $n$. This prediction contradicts  experimental observations.
 Such a situation could   result for two reasons:(1) the measured 
$p_T$ are larger
 than typical momentum $Q_0$, and this contribution is suppressed, as has
 been discussed in \eq{CORSUM}; and (2) the measurements were  not 
made in an
 inclusive type of the experiment, in which all events were summed without
 selection on multiplicities of the secondary hadron, but only events with
 large multiplicity were  measured. 
  \begin{boldmath}     
 \subsubsection{Measurements with fixed multiplicity $N = m \bar{n}$,}
 \end{boldmath}

\begin{figure}[ht]
\begin{tabular}{c c c}
 \includegraphics[width=6cm]{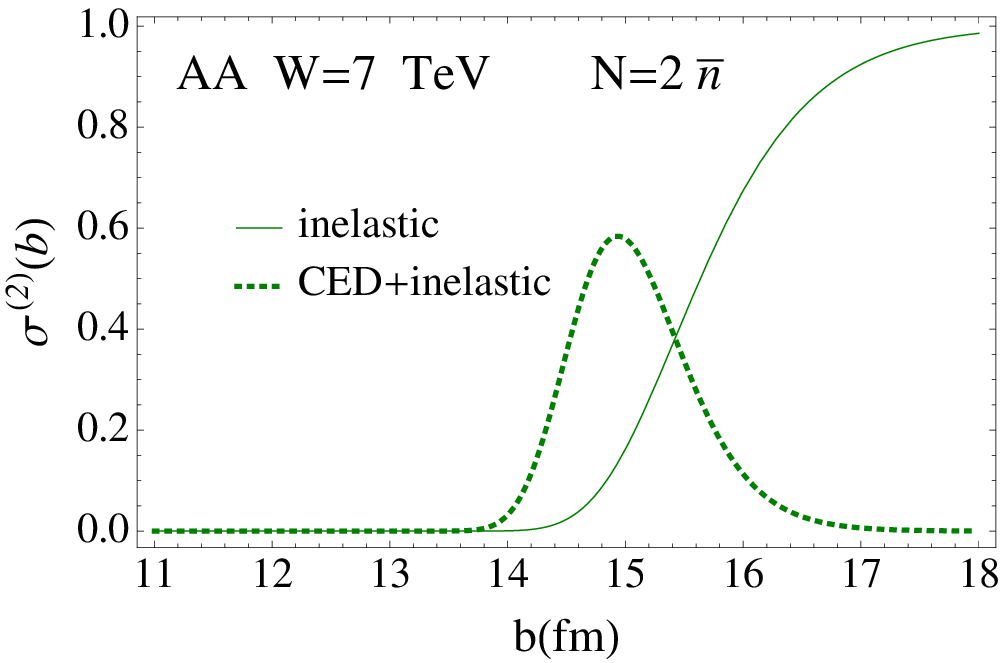}& \includegraphics[width=6cm]{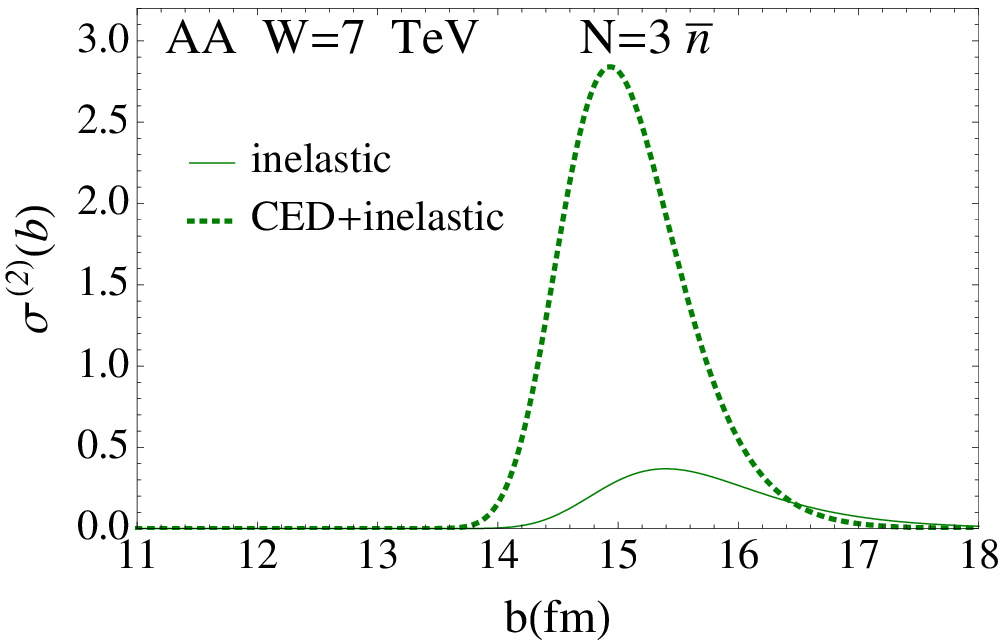}& \includegraphics[width=6cm]{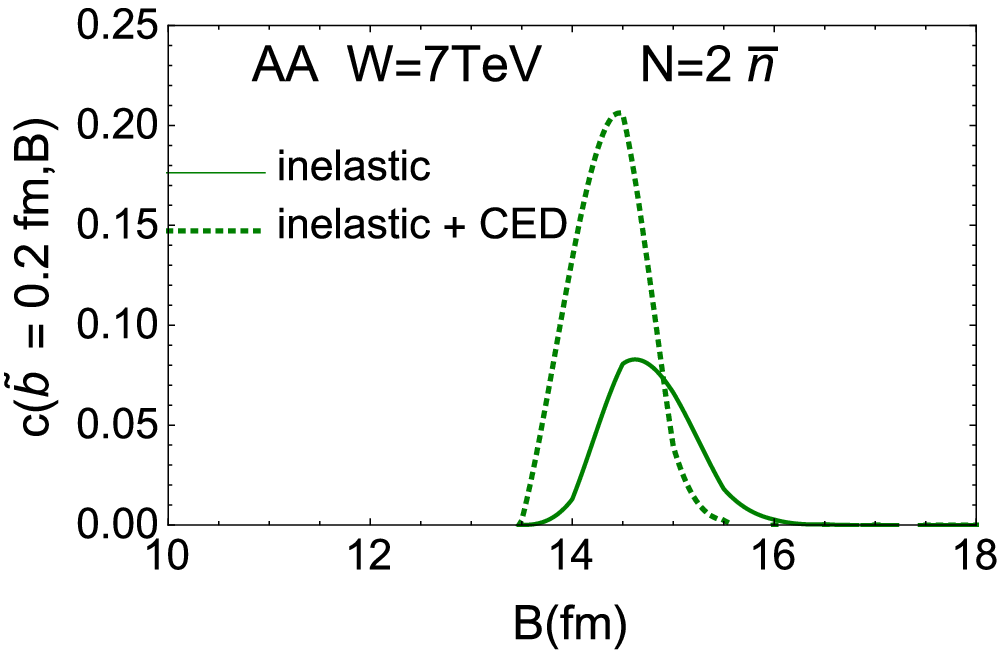}\\
 \fig{cedcom}-a &  \fig{cedcom}-b&  \fig{cedcom}-c\\
 \end{tabular}
    \protect\caption{Comparison of the inelastic events with the multiplicity
 $N = m \bar{n}$: for the  production of two identical gluons from
 the $m$-parton 
showers, and   central diffraction production in the event: 
 \fig{cedcom}-a for m =2 and \fig{cedcom}-b for $m =3$. \fig{cedcom}-c
 shows the same contribution as \fig{cedcom}-a  after all integrations.
.}
\label{cedcom}
   \end{figure}

First,  we would like to examine what happens to the
 symmetry   $ \varphi \to \pi - \varphi$  in  an event
 with given multiplicity.
We need to compare the production of $m$ parton showers which generate
 the event
 with multiplicity $N=m \bar{n}$, with the event with the same
 multiplicity, but in
 which we produce in addition the low multiplicity events, by  central 
diffraction
 production. From the point of view of the AGK cutting rules, the first 
process, is the
 process with $m$-cut Pomerons, while the second, is the process with the same
 $m$-cut Pomerons, plus two Pomerons which are not cut. At first sight, the
 second
 case could have a larger cross section, since it has an additional
 factor $ 
(\sigma_{\rm in}\,T_A\Lb b\Rb)^2$, which can be large for nucleus-nucleus 
scattering.  We need to estimate this contribution since it
 is suppressed
 by factor $\exp\Lb - 2 \Omega\Rb$ in \eq{NBE4}. In \fig{cedcom} we plot the
 $b$-dependence of $\sigma^{(m)}\Lb b\Rb$ of \eq{NBE4}, together with the 
coefficient
 from the AGK cutting rules.  From this figure we see that the
 processes of
 central diffraction in the inelastic environment is dominant, except for 
the 
process with $N = 2 \bar{n}$ which needs  additional consideration.
 This fact is a bit surprising since
 
\bea \label{NBE71}
\mbox{Inelastic production:} & N = m \bar{n}& \sigma^{(m)}_{in}\,\propto\,
\frac{m (m - 1)}{m!} \Lb 2 \Omega\Lb b \Rb\Rb^{m - 2} \,\exp\Lb - 2 \Omega\Lb
 b \Rb\Rb;\nn\\
\mbox{Inelastic production + CED:} & N = m \bar{n}& \sigma^{(m)}_{CED}\,\propto
\,\frac{2}{m!} \Lb 2 \Omega\Lb b \Rb\Rb^{m } \,\exp\Lb - 2
 \Omega\Lb b \Rb\Rb;
\eea 
The survival probability  $\exp\Lb - 2 \Omega\Lb b \Rb\Rb$, is very small 
at
 all $b$ less than $2 \,R_A$, and determines 
the  value  for
 $2 \Omega\Lb b \Rb \,\leq\,1$. Therefore, the extra  factor $\Lb 2
 \Omega\Rb^2$,  is not an enhancement, but a suppression (see 
\fig{cedcom}-c).   Nevertheless, it turns out that together with numerical coefficients this kind of suppression does not work.

However,
we
 need to consider the contribution to the correlation function, which 
includes
 the additional integrations over impact parameters,
\bea \label{corF} 
C\Lb \vec{p}_{1,T} \pm \vec{p}_{2,T}\Rb\,\,&=&\,\,\int d^2 \tilde{b} 
\,e^{( \vec{p}_{1,T}\, \pm\, \vec{p}_{2,T}) \cdot \vec{\tilde{b}}} 
\,\tilde{c}\Lb \vec{\tilde{b}}\Rb; ~~~~~~~~~~~~~~\tilde{c}\Lb \vec{\tilde{b}}\Rb\,=\,\int d^2 B\,c\Lb \vec{\tilde{b}}, \vec{B}\Rb \nn\\
c\Lb \vec{\tilde{b}}, \vec{B}\Rb\,\,&=&\,\,\int 
 d^2 b\,S_A\Lb \vec{B}  + \h \vec{\tilde{b}}\Rb\, S_A\Lb \vec{B}  - \h
 \vec{\tilde{b}}\Rb \, S_A\Lb \vec{b}  + \h \vec{\tilde{b}}\Rb\, S_A\Lb \vec{b}
  - \h \vec{\tilde{b}}\Rb\, \sigma^{(m)}_{in,CED}\Lb \vec{B}\Rb
  \eea
    
Integration over all impact parameters shows that in  the event with
 $N = 2 \bar{n}$,  the process with dijet production is also larger than
 the Bose-Einstein correlations (see \fig{cedcom}-c).

 One can see that the multiparticle production accompanied by exclusive
 production of two gluon jet prevails, leading to negative $v_{n,n}$,
 for odd $n$.    For nucleus-nucleus  collisions, it is well known, 
that
 this statement contradicts the experimental data 
\cite{CMSAA,ALICEAA,ATLASAA}.

  \begin{boldmath}     
 \subsubsection{Measurements with multiplicity $N \geq  m \bar{n}$,}
 \end{boldmath}
 Hence, for nucleus-nucleus scattering, the inclusive experiments, as
 well as the measurements with fixed multiplicity in the Leading Log(1/x)
 Approximation of perturbative QCD, generate negative $v_{n,n}$ for odd
 $n$, which contradicts the experimental data.   In this subsection 
 we examine the situation when the events with multiplicities larger
 that $m_0 \bar{n}$ ($N \geq m_0 \bar{n}$) is measured, as it has been
 done in the most experiments. Summing \eq{NBE71} over all $m \geq m_0$
 we obtain
\bea 
\sigma^{m_0}_{\rm in}\Lb Y; B \Rb \,\,&=&\,\,1 \,\,-
\,\,\frac{\Gamma\Lb m_0 - 2, 2  \Omega\Lb B; Y\Rb \Rb}{\Gamma\Lb m_0
 - 2\Rb}\,\,\,\,\xrightarrow{\Omega \gg 1} \,\,\,\,1 -   \frac{\Lb 2
 \Omega\Lb B; Y\Rb\Rb^{m_0-3}}{(m_0 - 3)!} e^{ - 2\Omega\Lb B; Y\Rb}
 ;\label{cor1}\\
\sigma^{m_0}_{\rm CED}\Lb Y; B \Rb &=& 2 \Big( 1 - \frac{\Gamma\Lb m_0,
 2  \Omega\Lb B; Y\Rb\Rb}{\Gamma\Lb m_0\Rb}\Big)\,\,\,\,\,\,\,\,\,
\,\xrightarrow{\Omega \gg 1} \,\,\,\,2\Big( 1 -   \frac{\Lb 2 \Omega\Lb B;
 Y\Rb\Rb^{m_0-1}}{(m_0 - 1)!} e^{ - 2\Omega\Lb B; Y\Rb}\Big) ;\label{cor2}
\eea

\begin{figure}[ht]
\begin{tabular}{c c c }
  \includegraphics[width=7.5cm]{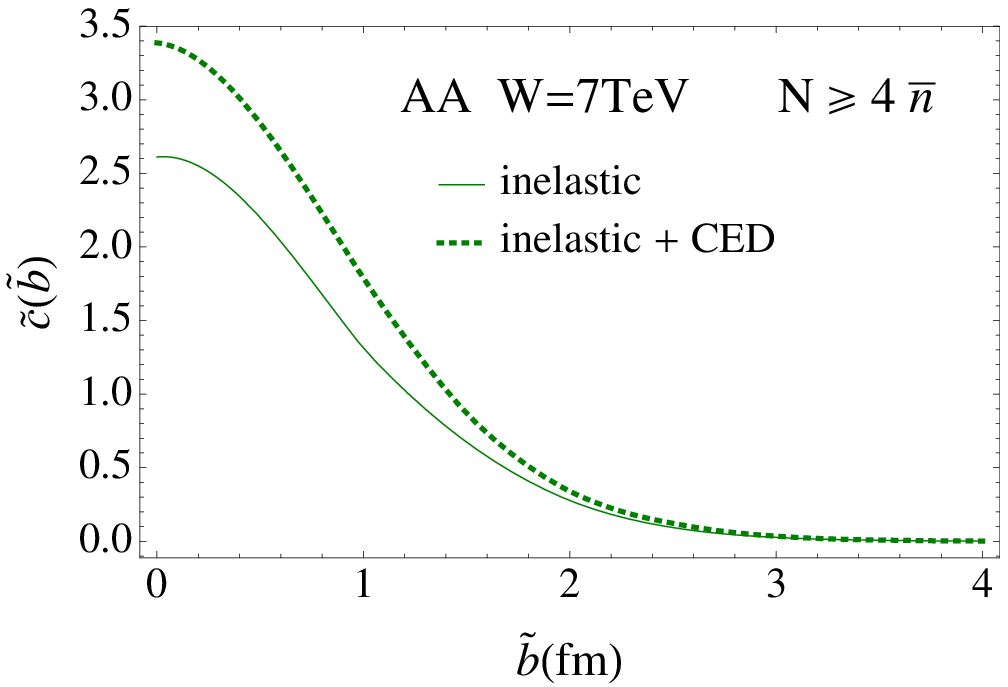}&~&\includegraphics[width=7.65cm]{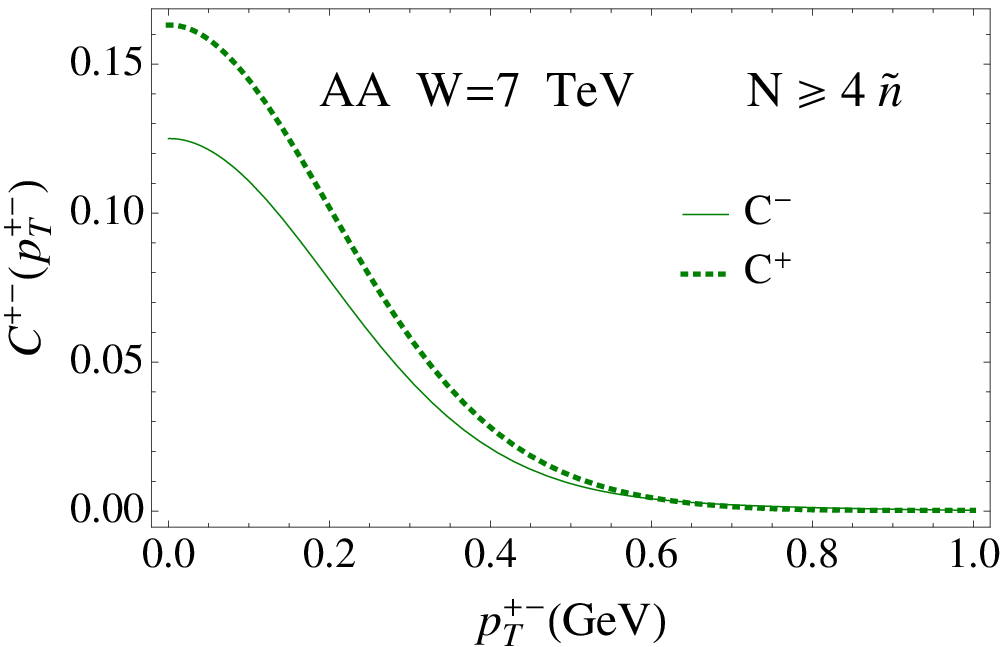}\\
 \fig{xsB}-a &~~~~~&  \fig{xsB}-b\\
 \end{tabular}
    \protect\caption{Comparison of the inelastic events with the multiplicity
 $N \geq m_0 \bar{n}$ in gold-gold collision at $W=7 \,TeV$: for the
  production of two identical gluons for larger than $m_0$-parton 
showers, and   central diffraction production in the event with the
 multiplicity not smaller than $m_0 \bar{n}$ .  \fig{xsB}-a shows the
 contribution of inelastic event and inelastic even plus central
 diffraction, for $m_0 =4$. In \fig{xsB}-b we plot the correlation
 functions $C^{\pm}\Lb | \vec{p}_{1,T} \pm \vec{p}_{2,T}|\Rb$
 (see \eq{BEDD2} and \eq{BEDD3}). $ p^{\pm}_T\, \equiv\, | \vec{p}_{1,T}
 \pm \vec{p}_{2,T}|$.}
\label{xsB}
   \end{figure}

One can see that at large $\Omega$, that the inelastic event with 
additional 
dijet
 production, is larger that the inelastic event that generates the 
Bose-Einstein correlations.
    In \fig{xsB}-a we plot the  function $\tilde{c}\Lb\tilde{b}\Rb$ of 
\eq{corF},
 which also shows that the inelastic contribution with  dijet 
production
 prevails. \fig{xsB}-b shows the correlation functions of \eq{BEDD2} and
 \eq{BEDD3}. Note that the Bose-Einstein correlations are smaller than
 the correlations due to the diffractive production of dijets.

Hence,  the experimental results are in direct
 contradiction with the theoretical predictions based on
 the Leading Log(1/x) Approximation of perturbative QCD.
 The only explanation that we can suggest  is, that the Sudakov  
form 
 factor suppresses the dijets production.

We believe that the p measured $p_T$ turns out to be much larger than 
$Q_0$, and
 double log suppression results in a small contribution of the process of
 central diffraction.   Indeed, for the exchange of the BFKL
 Pomeron our value for $Q_0 \approx Q_s\Lb y_1\Rb$  appears  to be
 overestimated. Our conclusions that typical $k_T \approx Q_s$ is
 based on the diagrams of \fig{froi}-a and \fig{froi}-b, in which 
the same diagrams contribute  to central diffraction and 
the inclusive
 cross section. However, for the exclusive central production there
 no AGK cutting rules, and the diagrams of \fig{froi}-c  should be taken
 into account.  If we  remove  the integral in \eq{VG5} for the 
Sudakov form
 factor, the  remaining expression takes the form of \eq{VG6}. For 
the BFKL Pomeron, it
 is just the contribution to the total cross section.  The typical
 transverse momenta in the BFKL Pomeron, both increase and decrease
 as function of rapidity (see Ref\cite{BARSIG}) and at large $y_2$
 or $Y - y_1$, the typical $k_T$ is as small as the non-pertutbative
 soft momentum,  which could be of the order of $ \Lambda_{\rm QCD}$.
 If we replace the emission of gluons by
 \eq{VG6}, 
 the diagrams of Fig.5-c
  reduce to the contribution to the total cross section, 
supporting the idea that $Q_0$ is of the order of typical soft momentum.
 Therefore, we expect that $Q_0 \approx \mu_{\rm \tiny soft}
 \approx \Lambda_{QCD}$.
 Bearing this in mind we concentrate our efforts below on the
 calculating  Bose-Einstein correlations, and their dependence
 on multiplicity of the events.

 ~

 
 \section{Dependence of  Bose-Einstein correlations  on the
 multiplicity of the event}


   In this section, we  consider  the dependence of
  Bose-Einstein correlations
 on the multiplicity of the event, using the Glauber formula for the
 total cross
 section. In accord with the AGK cutting rules, the multiplicity of
 the event ($N$)
 is intimately related to the number of parton showers ($ m $) that
 are produced,
 where $N = m\,\bar{n}$.  
   
   In the framework of this approach, the Bose-Einstein correlations
 in the event with
 multiplicity $N\, = \,m \,\bar{n}$  is determined by the following
 expression (see also \eq{corF}):
 \bea
 \frac{d^2 \sigma}{d y_1 \,d y_2  d^2 p_{T1} d^2 p_{T2}  }\,\,&\propto&\,\, C_A\Lb L_c |\vec{p}_{12,T}|\Rb\,\frac{d \sigma^{\rm BFKL}}{d y_1 \,d^2 p_{T1} } \,\frac{d \sigma^{\rm BFKL}}{d y_2 \,d^2 p_{T2} }; \\
C_A\Lb L_c |\vec{p}_{12,T}|\Rb& =&\frac{1}{N^2_c - 1} \frac{I\Lb L_c |\vec{p}_{12,T}|\Rb}{I\Lb 0\Rb},~~~~~ I\Lb L_c |\vec{p}_{12,T}|\Rb=\int\!\! d^2 \tilde{b}\,e^{ i \vec{\tilde{b}}\cdot \vec{p}_{12,T}}\,{\cal I}\Lb \tilde{b} \Rb\label{NBE2}\\
 {\cal I}\Lb \tilde{b} \Rb\,&=&\,\int d^2  B  \,c\Lb \vec{\tilde{b}},\vec{B}\Rb \label{NBE3}\\
 \sigma^{(m)}\Lb \vec{B} + \vec{b}\Rb& = &\sum^\infty_{n=m, m \geq 2} \,m\,\Lb m - 1\Rb \frac{\sigma^{(m)}_n}{\Omega^2\Lb s, \vec{B}\Rb}\,\,=\,\,\frac{\Lb 2 \Omega\Lb s, \vec{B} \Rb\Rb^{m-2}}{(m - 2)!} e^{-2 \Omega\Lb s, \vec{B} \Rb} \label{NBE4} \eea   
 
\begin{figure}[ht]
\begin{tabular}{c c c}
 \includegraphics[width=7cm]{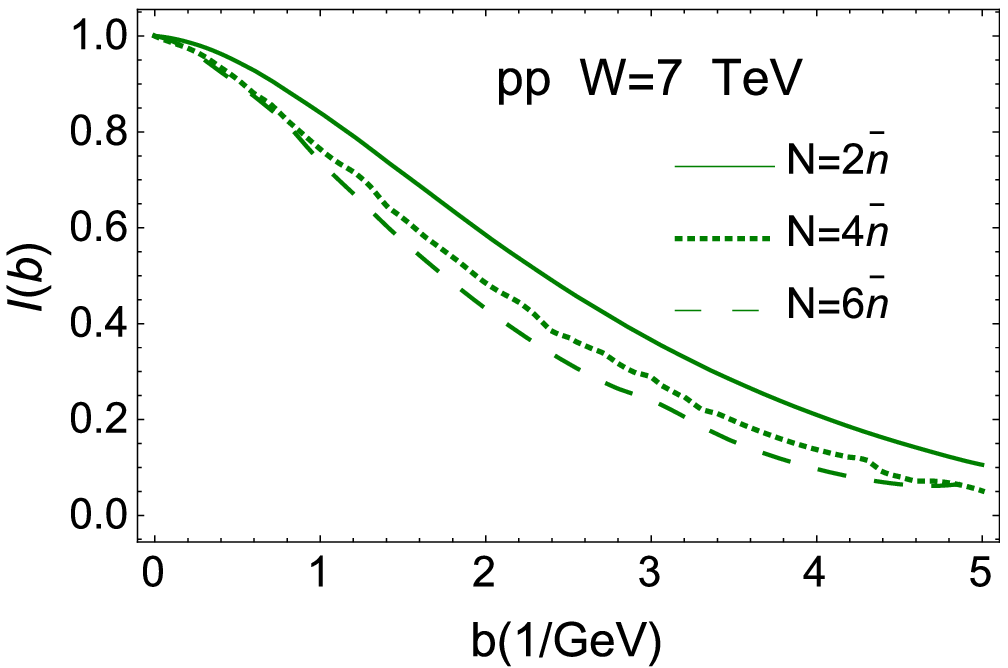}& \,\,\,\,\,\,\,\,\,\,
\,\,& \includegraphics[width=7cm]{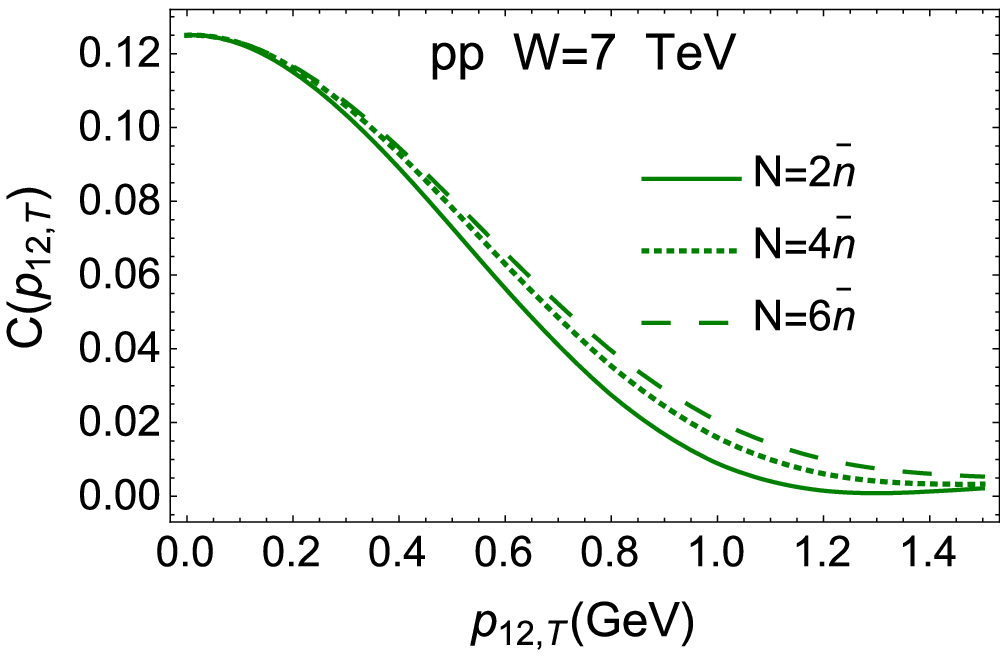}\\
 \fig{cornpp}-a & ~~~~~   & \fig{cornpp}-b\\
 \end{tabular}
    \protect\caption{\fig{cornpp}-a shows $ {\cal I}\Lb b \Rb$ for
 proton-proton
 scattering with the parameters, that are given in \eq{NBE6},  as a
 function of
 $b$, for the events with different multiplicities normalized to 1
 at $b = 0$. In
 \fig{cornpp}-b the correlation function $C\Lb p_{12,T}\Rb$ is plotted
 versus
 $p_{12,T}$. $\bar{n}$ is the average multiplicity in the single
 inclusive production.}
\label{cornpp}
   \end{figure}

   If we assume $S_A\Lb b \Rb$  to have a Gaussian form i.e. $S_A\Lb 
b 
\Rb= \Lb A/\Lb \pi
 R^2_A\Rb\Rb\exp\Lb - b^2/R^2_A\Rb$, then \eq{NBE3} takes the form
    \beq \label{NBE5}
 {\cal I }\Lb \tilde{b}\Rb =  \Lb\frac{A}{\pi \,R^2_A} \Rb^4\,e^{- 
\frac{\tilde{b}^2}{R^2_A}}\Bigg( \int\!\!d^2B\, d^2 b \,e^{ -2 \frac{(B^2
 + b^2)}{R^2_A}}\, \frac{\sigma^{(m)}_n\Lb \vec{B} + \vec{b}\Rb}{\Omega^2\Lb s,
 \vec{B} + \vec{b}\Rb}\Bigg)
 \eeq   
  
  and the correlation function does not depend on $m$ or, in other words,
 it does not depend on the multiplicity of the event.  However, this result
 is the specific property of the Gaussian approximation, which cannot be correct
 even for hadron-hadron collisions, since it does not lead to the  correct 
exponential
 behaviour of the scattering amplitude at large impact parameters $b$. 
  Considering
 the Glauber model for the description of the proton-proton scattering at high
 energies, we replace $S_A $ and $T_A$ in \eq{CA} and \eq{CAB} by 
  \beq \label{NBE6}
  S_N\Lb b \Rb\,\,=\,\,\frac{m^2}{2 \,\pi} K_0\Lb m b \Rb;~~~~~~~~~ T_N\,=\,\int
 d^2 b' S_N\Lb b'\Rb\,S_N\Lb \vec{b} - \vec{b}'\Rb;~~~~~~\Omega\,=\,\sigma_0
 \,e^{\Delta\,Y} T_N\Lb b \Rb
  \eeq
  where $\sigma_0 = 4\, 1/GeV^2$, $m = 1 \,GeV$ and $\Delta = 0.1$, were 
chosen
 to describe the value and  energy behaviour of the total cross section for
 the proton-proton interaction at high energy. In \fig{cornpp}-a the behaviour
 of ${\cal I}\Lb b \Rb$ is shown for the events with different multiplicities.
 We see that the correlation length $L_c$ decreases as function of the
 multiplicity. In other words, the typical momentum in the correlation function
 $C\Lb L_c p_{12,T}\Rb$  increases with $N$, as  can be seen from 
\fig{cornpp}-b,
 where the value of the correlation function $C\Lb L_c p_{12,T}\Rb$ is plotted.
 
\begin{figure}[ht]
\begin{tabular}{c c c}
   \includegraphics[width=7.3cm]{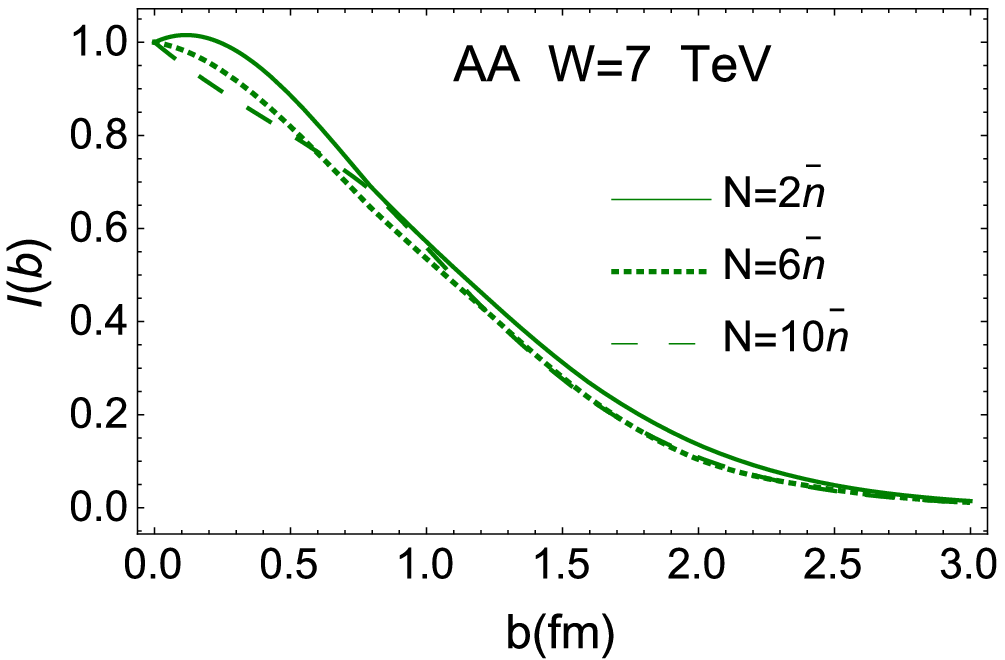}&~~~& \includegraphics[width=7.5cm]{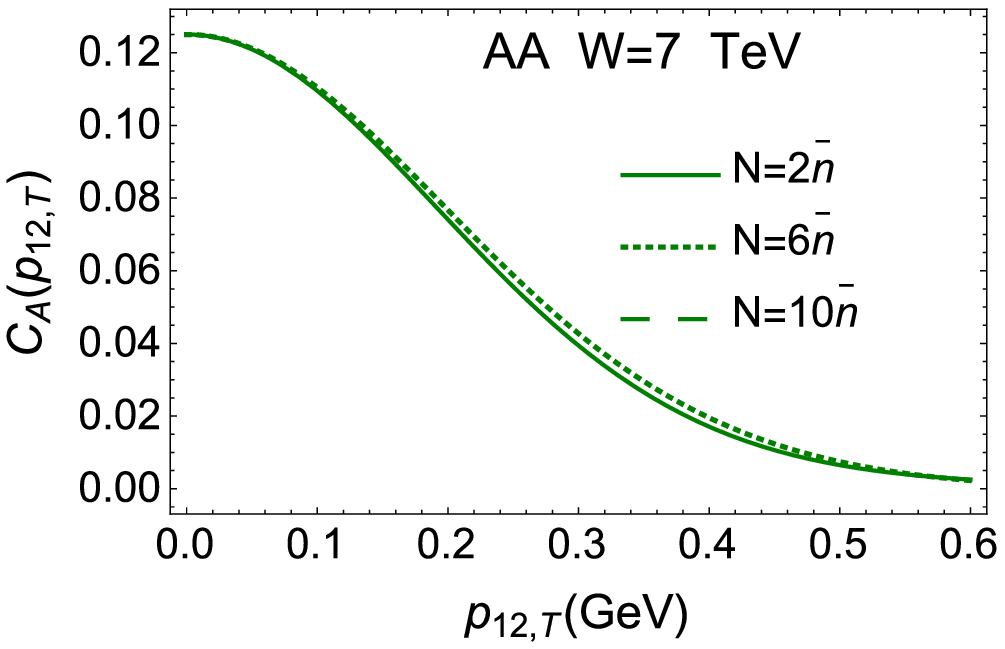}\\
 \fig{cornaa}-a &   &\fig{cornaa}-b\\
 \end{tabular}
    \protect\caption{$ {\cal I}\Lb b \Rb$ for nucleus-nucleus (gold-gold)
 scattering
 with  $S_A\Lb b \Rb$  given in \eq{CA}, as a function of $b$, for 
the 
events
 with different multiplicities. In \fig{cornaa}-a  ${\cal I}\Lb b \Rb$ are 
normalized to
 their values at $b=0$. $\bar{n}$ is the average multiplicity in the
 single inclusive
 production. The correlation function $C\Lb p_{212,T}\Rb$ is plotted
 in \fig{cornaa}-b.}
\label{cornaa}
   \end{figure}

 The correlation length of  the correlation function in nucleus-nucleus 
collisions, shows only
 mild dependence on the multiplicity of the events, (see \fig{cornaa} -b,
 while the
 value of ${\cal I} $ crucially depends on $N$ (see \fig{cornaa}-a).
 \fig{cornaa}-c
 shows that the correlation function $C_A\Lb L_c p_{12,T}\Rb$ does not
 depend on the
 multiplicity of the event.\\

 {}

For completeness of presentation we calculated both ${\cal I}\Lb b\Rb$
 and $C_{pA}\Lb p_{12,T}\Rb$ for proton-gold scattering.

\begin{figure}[ht]
\begin{tabular}{c c c}
 \includegraphics[width=7cm]{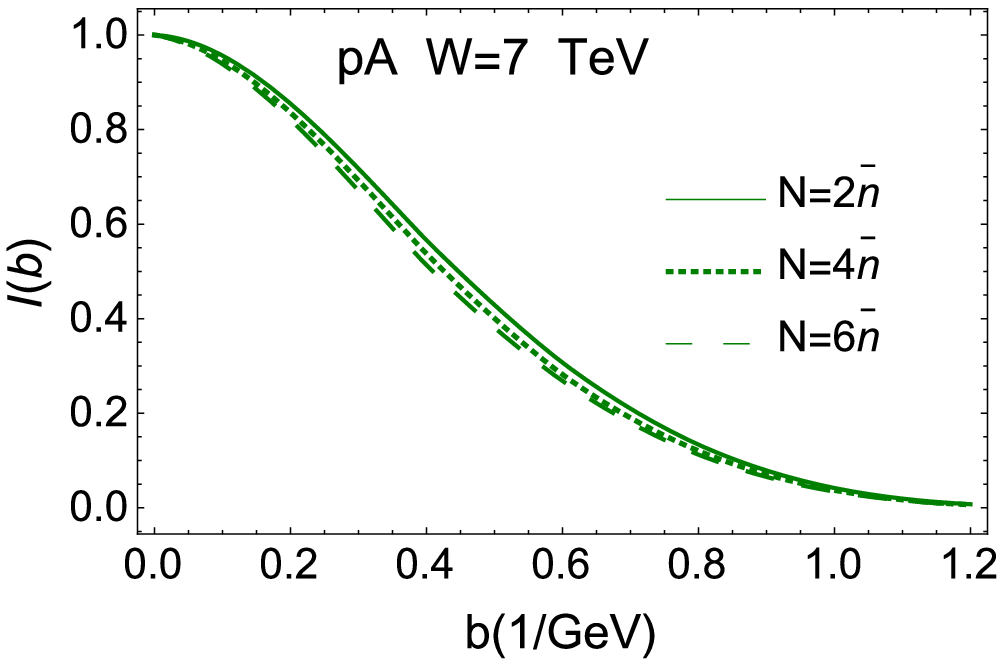}& \,\,\,\,\,\,\,\,\,\,\,\,& \includegraphics[width=7cm]{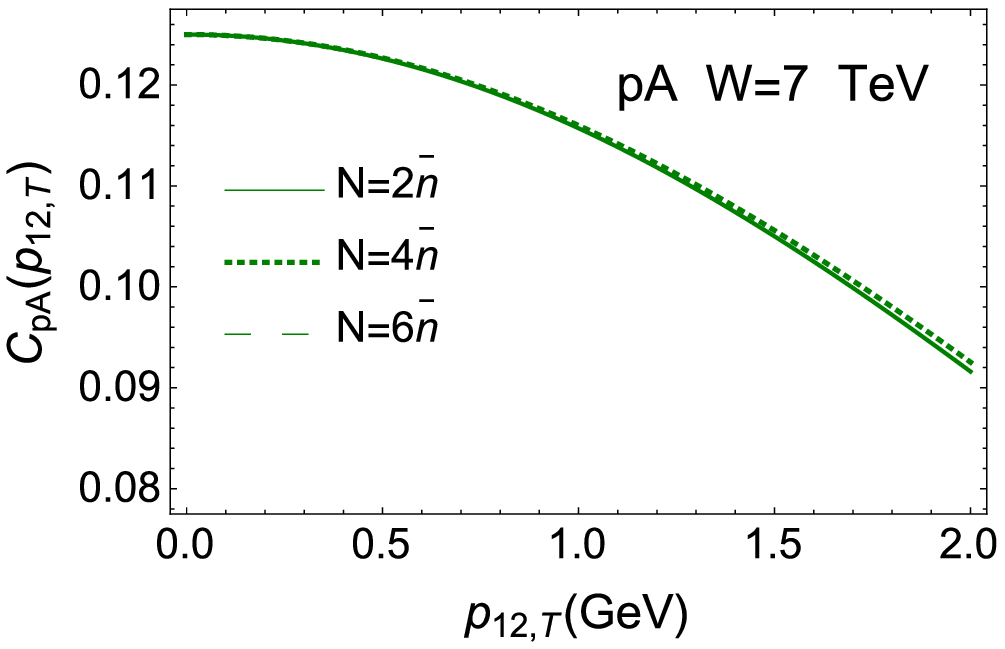}\\
 \fig{cornpa}-a & & \fig{cornpa}-b\\
 \end{tabular}
    \protect\caption{\fig{cornpa}-a shows $ {\cal I}\Lb b \Rb$ for proton-gold 
 scattering with the parameters  that are  given  in \eq{CA}, and with 
the
 typical $b= 1\,1/GeV$  in  proton-proton scattering,  as a function of
 $b$, for the events with different multiplicities, normalized to 1 at $b = 0$.
 In \fig{cornpa}-b the correlation function $C_{pA}\Lb p_{12,T}\Rb$ is
 plotted versus $p_{12,T}$. $\bar{n}$ denotes
 the average multiplicity in  single inclusive production.}
\label{cornpa}
   \end{figure}

The results of these calculations are plotted in \fig{cornpa}. The first
 observation is that the correlation length does not depend on the 
size of
 the nucleus, and is determined by the typical impact parameter
 in proton-proton
 scattering. The dependence on multiplicity of the event is rather mild.

Concluding this section, we would like to emphasis that the dependence on 
multiplicity
 due to the  production of several parton showers, turns out to be mild,
 except for the 
case 
of hadron-hadron collisions. For this collision the larger multiplicity of the 
event,
 the shorter is the correlation length $L_c$,
or, in other words, the typical momentum increases in the events with large
 multiplicities. On the other hand, such an increase is not very pronounced,
 and even for hadron-hadron collisions, we can expect that the main
 source of the 
multiplicity
 dependence is from  the structure of one parton shower.  In the
 next section,
 we  discuss
 the saturation of the parton density in the one parton shower
  for nucleus-nucleus collisions, and we 
 develop a simple model in the spirit of the KLN approach.

{}

\section{ A simple KLN -type model for the structure of one parton cascade 
in CGC}


\subsection{ Momentum dependence    of the BFKL Pomeron in a 
nucleus.}

\begin{figure}[ht]
 \includegraphics[width=13cm]{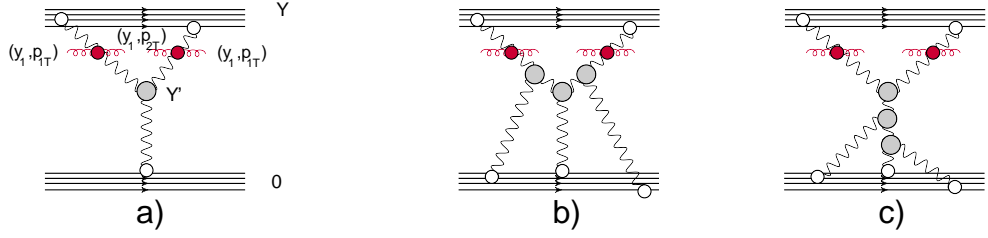}
   \protect\caption{ The double inclusive production for ion-ion 
collisions which
 lead to the azimuthal correlations with the correlation length
 $L_c \propto 1/Q_s$:
  the first diagram is displayed in \fig{enh}-a, while \fig{enh}-b shows the
 interaction of the BFKL Pomerons which results in $L_c \propto 
1/Q_{s,A}$,
 where $Q_{s,A}$ denotes the saturation momentum of the nucleus.
 The wavy lines denote the BFKL Pomerons. The  red blobs show the Mueller
 vertices for two gluons production, while the gray circles stand for the
 triple Pomeron vertices.   The white circles show the vertex of the interaction
 of the BFKL Pomeron with the nucleon in the nucleus.The produced  gluons are
 denoted  by  red helical lines. For simplicity we draw the diagrams 
at $y_1=y_2$.}
\label{enh}
   \end{figure}

As we have seen, the diagrams in which the  structure of the one parton 
shower 
is
 described by the BFKL Pomeron, lead to the  correlation length of  azimuthal
 angle correlations $L_c \propto 1/R_A$ or, in other words, to the typical
 transverse
 momentum which is very small (see \fig{cornaa}). Therefore,  we need to
 discuss a more
 complicated structure of the single parton shower, which is related, for
 example,
 to `fan' diagrams shown in \fig{froi}-b. We expect that the interaction 
of the 
BFKL
 Pomeron will lead to  the value of $L_c \sim 1/Q_{s,A}$, where $Q_{s,A}$
 denotes 
the
 nucleus saturation momentum. In particular, we consider the diagrams of
 \fig{enh}-a
 and \fig{enh}-b. The diagram of \fig{enh}-a is the first diagram 
that
 leads to the
 correlation function  which depends on the saturation momentum of 
the
 nucleon, as  
 shown in Ref.\cite{GLMBE,GOLE}. We will show that the interaction of
 the BFKL Pomerons
 with the nucleus,  examples of which are shown in \fig{enh}-b, will lead to 
$L_c 
\propto 1/Q_{s,A}$.

\begin{figure}[ht]
 \includegraphics[width=16cm]{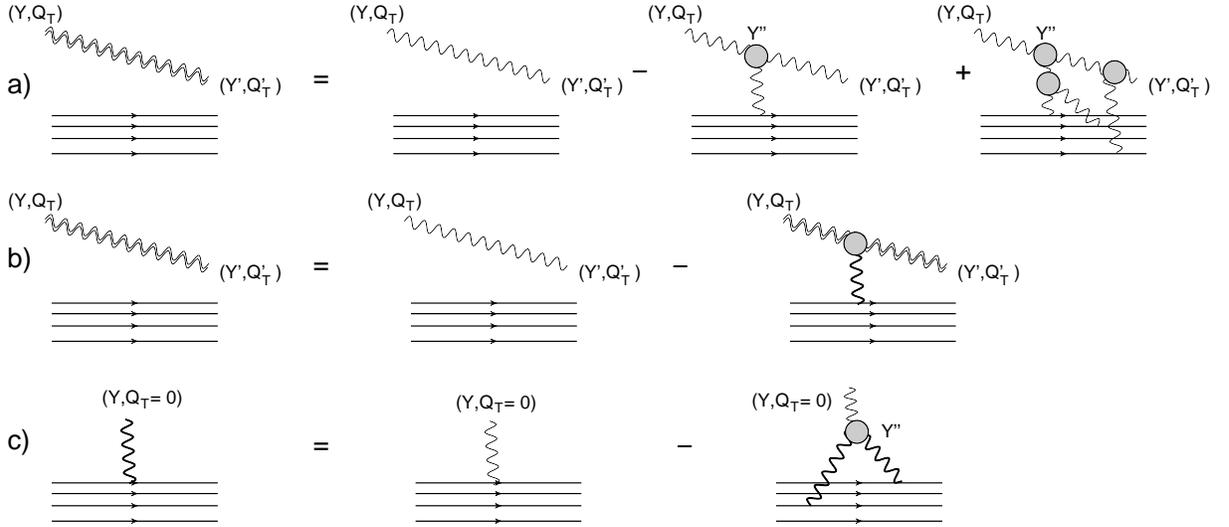}
   \protect\caption{ Equations for BFKL Pomeron propagator in the
 nucleus. \fig{eq}-a shows the first simple diagrams. \fig{eq}-b presents
 the equation for the propagator. \fig{eq}-c describes the Balitsky-Kovchegov
 equation. Wavy lines describes the BFKL Pomerons. The double wavy lines denote
 the resulting propagator. The bold wavy lines stand for the solution of
 Balitsky-Kovchegov equation in the nucleus. The blobs denote
 the triple Pomeron vertices.}
\label{eq}
   \end{figure}

The general equation for the propagator of the BFKL Pomeron in a nucleus 
is
 shown in \fig{eq}. The simplest form these equation have in the framework
 of Gribov Pomeron Calculus \cite{GRIB} with $\alpha'_\pom=0$ and the 
Pomeron
 intercept $\Delta$.
Denoting by $T_A\Lb Y, Q_T; Y' Q'_T\Rb$ and $G_A\Lb Y,Q_T\Rb$ the
 dressed (resulting) propagator of the Pomeron, and the solution of
 the non-linear Balitsky-Kovchegov equation of \fig{eq}-c , respectively,
 the equations take the forms
\bea 
&&\hspace{-0.7cm}T_A\Lb Y, Q_T; Y' Q'_T\Rb = T\Lb Y-Y',Q_T\Rb -\Delta\,\int^Y_0\!\!\!\!\! dY'' d^2 Q''_T \,T\Lb Y-Y'',Q_T\Rb\,G_A\Lb Y'', \vec{Q}_T - \vec{Q}''_T\Rb\,T_A\Lb Y'', Q''_T; Y' Q'_T\Rb;\label{EQ10}\\
&&\hspace{-0.7cm}\frac{\partial T_A\Lb Y, Q_T; Y' Q'_T\Rb}{\partial Y} = \Delta\Bigg(  T_A\Lb Y, Q_T; Y' Q'_T\Rb\,\,-\,\, \int\!\!d^2Q''_T\, G_A\Lb Y, \vec{Q}_T - \vec{Q}''_T\Rb\,T_A\Lb Y, Q''_T; Y' Q'_T\Rb\Bigg)\label{EQ11}; \\  
&&\hspace{-0.7cm}T\Lb Y-Y',Q_T\Rb  =  g\Lb Q_T\Rb \exp\Lb \Delta \Lb Y - Y'\Rb\Rb; ~~~~~~~~~~~~T_A\Lb Y=Y', Q_T; Y' Q'_T\Rb\,=\,g\Lb Q_T\Rb;\label{EQ12}\eea
 
\bea 
&&\hspace{-0.7cm}G_A\Lb Y, Q_T\Rb = G\Lb Y-Y',Q_T\Rb -\Delta\,\int^Y_0dY'' d^2 Q''_T \,G^0\Lb Y-Y'',Q_T\Rb\,G_A\Lb Y'', \vec{Q}_T - \vec{Q}''_T\Rb\,G_A\Lb Y'', Q''_T; Y' Q'_T\Rb;\label{EQ20}\\
&&\hspace{-0.7cm}\frac{\partial G_A\Lb Y, Q_T\Rb}{\partial Y} = \Delta\Bigg(  G_A\Lb Y, Q_T\Rb\,\,-\,\, \int d^2Q''_T\, G_A\Lb Y, \vec{Q}_T - \vec{Q}''_T\Rb\,G_A\Lb Y, Q''_T\Rb\Bigg)\label{EQ21}; \\  
&&\hspace{-0.7cm}G^0\Lb Y-Y',Q_T\Rb  =   \exp\Lb \Delta \Lb Y - Y'\Rb\Rb; ~G_A\Lb Y=0,Q_T\Rb\,=\,S_A\Lb Q_T\Rb~\mbox{with}~S_A\Lb Q_T\Rb = \int d^2 b e^{i \vec{Q}_T \cdot \vec{b}}S_A\Lb b \Rb;\label{EQ22}\eea

  The main idea of solution, is the observation that in $G_A\Lb Y,Q_T\Rb $
 the typical 
$Q_T \sim 1/R_A\,\ll\,1/R_N$ or $Q_s$, where $R_N$ is the nucleon size.
 Therefore, in
  \eq{EQ10}-\eq{EQ22}  we can replace
 $G_A\Lb Y,Q_T\Rb$ by $\int d^2 Q_T \,G_A\Lb Y,
 Q_T\Rb \delta^2\Lb Q_T\Rb$. At $Y=0$,  $\int d^2 Q_T\,
 G_A\Lb Y, Q_T\Rb = S_A\Lb
 b=0\Rb \propto 2 \rho R_A$, where $\rho$ denotes the density of the 
nucleons
 in a nucleus.  
  Plugging this expression in the above equations,  they reduce  
to
 the following form
  \bea 
  \frac{d  T_A\Lb Y, Q_T; Y' Q'_T\Rb}{d Y} &= &\Delta\Bigg(  T_A\Lb Y, Q_T; Y'
 Q'_T\Rb\,\,-\,\, \,\tilde{G}_A\Lb Y\Rb\,T_A\Lb Y, Q_T; Y' Q'_T\Rb\Bigg);\label{EQ31}\\
\frac{d  \tilde{G}_A\Lb Y \Rb}{d Y} &=& \Delta\Bigg(  \tilde{G}_A\Lb Y\Rb\,\,-
\,\,\,\,\tilde{G}^2_A\Lb Y\Rb\Bigg)~ \mbox{where}~ \tilde{G}_A\Lb Y\Rb =\int d^2
 Q_T\, G_A\Lb Y, Q_T\Rb\label{EQ32}
\eea

Solving \eq{EQ32} and \eq{EQ31} we obtain
\beq \label{EQ4}
\tilde{G}_A\Lb Y\Rb\,=\,\,\frac{S_A\Lb b = 0\Rb e^{\Delta Y}}{1 +
 S_A\Lb b = 0\Rb \Lb e^{\Delta Y} - 1\Rb};~~~~~~~T_A\Lb Y, Q_T; Y'
 Q_T\Rb\,\,=\,\,g\Lb Q_T\Rb\,e^{\Delta\Lb Y - Y'\Rb} \frac{ 1\,+\,
 S_A\Lb b = 0\Rb\, \Lb e^{\Delta Y'} - 1\Rb}{ 1\,+\, S_A\Lb b = 0\Rb\,
 \Lb e^{\Delta Y} - 1\Rb};\eeq

In  the   general case, the equations have a more complicated 
structure, and include
 the dependence on the transverse momenta, which are the  Fourier images 
of
 the dipole sizes. However, in the vicinity of the saturation scale, the
 scattering amplitude displays a geometric scaling behaviour\cite{GS},  and
  depends only on one variable $Q^2_s/p^2_T$.  In the vicinity of 
the
 saturation scale  the equations take the form:

  \bea \label{EQ5}
  \frac{d  T_A\Lb z;z'\Rb}{d Y} &= &\Lb 1 - \gamma_{cr}\Rb\,\Bigg(
  T_A\Lb z, z'\Rb\,\,-\,\, \,\tilde{G}_A\Lb z \Rb\,T_A\Lb z, z'\Rb\Bigg);
\label{EQ51}\\
\frac{d  \tilde{G}_A\Lb z \Rb}{d z} &=&\Lb 1 - \gamma_{cr}\Rb\Bigg(  \tilde{G}_A\Lb z\Rb\,\,-\,\,\,\,\tilde{G}^2_A\Lb z\Rb\Bigg)\label{EQ52}\eea

Solutions of these equations have the following form:
\beq \label{EQ6}
\tilde{G}_A\Lb z\Rb\,=\,\,\frac{\phi_0\, e^{\Lb 1 - \gamma_{cr}\Rb \,z}}{1 +  \phi_0  \Lb e^{\Lb 1 - \gamma_{cr}\Rb\, z} - 1\Rb};~~~~~~T_A\Lb z, z'\Rb\,\,=\,\,g\Lb Q_T\Rb\,e^{\Lb 1 - \gamma_{cr}\Rb \Lb z - z' \Rb} \frac{ 1\,+\, \phi_0\, \Lb e^{ \Lb 1 - \gamma_{cr}\Rb\, z' } - 1\Rb}{ 1\,+\, \phi_0 \, \Lb e^{\Lb 1 - \gamma_{cr}\Rb\,z } - 1\Rb};
\eeq
where $\phi_0$ denotes the value of the scattering amplitude at $z=0$ and 
\beq  \label{EQ7}
z \,=\,\ln \Lb \frac{Q^2_{s,A}\Lb Y \Rb}{p^2_T}\Rb~~~\mbox{with}~~~~~
~Q^2_{s,A}\Lb Y \Rb \,\,=\,\,S_A\Lb b = 0\Rb\,Q^2_s\Lb Y  \Rb
\eeq
where $Q_S\Lb Y\Rb$ denotes the proton saturation momentum.

The principle feature of all these  solutions is that, the interaction
 with nucleus, which is shown in \fig{enh}-b and in \fig{eq}, does not
 affect the dependence on $Q_T$, which determines the angular 
correlations.
 The only diagrams that could depend on the nuclear saturation momentum,
 are shown in \fig{enh}-c. Generally speaking the BFKL Pomeron from
 the rapidity 0 to rapidity $Y'$, should be replaced by the dressed BFKL
 Pomeron (see \fig{incl}).

\subsection{The model.}


\subsubsection{The general formulae.}

\begin{figure}[ht]
 \includegraphics[width=5cm]{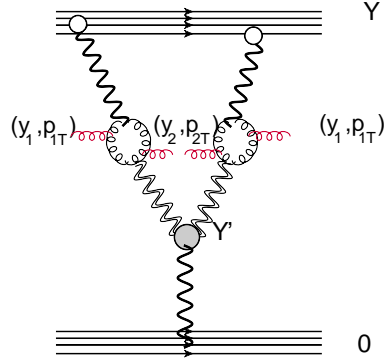}
   \protect\caption{ Double inclusive cross section.   The double wavy lines
 denote the propagator of the dressed BFKL Pomeron. The bold wavy lines stand
 for the solution of Balitsky-Kovchegov equation in the nucleus. Helical line
 denote gluons. }
\label{incl}
   \end{figure}

The diagram for the interference of two parton showers is shown in 
\fig{incl},
 and can be written in the form:
\bea \label{M1}
&&\frac{d^2 \sigma^{\rm interference\,diagram}}{d y_1 \,d y_2\,d^2 p_{1,T}\,d^2 p_{2,T}}\,\,\propto \\
&&\frac{\bas^2  V^2\Lb p_{1,T}, p_{2,T} , y_1 - y_2\Rb}{p^2_{1,T}\,p^2_{2,T}}\!\!\int^{z_1 \approx z_2}_0\!\!\!\!\!\! \!\!\!\!\!\! \!\!\!d z'
G_A\Lb z_Y\,-\, z_1 \Rb \,G_A\Lb  z_Y\, -\, z_2 \Rb \,T_A\Lb z_1\, -\, z' \Rb \,T_A\Lb  z_2\, - \, z'\Rb \,\Gamma_{3 \pom}\Lb  Q_T; Q_{s,A}\Lb Y'\Rb\Rb \,G_A\Lb z' \Rb\nn
\eea
Assuming $\bas\Lb y_1 \,-\,y_2\Rb\,\ll\,1$,  \,$V\Lb p_{1,T}, p_{2,T} ,
 y_1 - y_2\Rb$  takes the simple form
\beq \label{M2}
V\Lb p_{1,T}, p_{2,T} , y_1 - y_2\Rb\,\,=\,\,\Gamma_\mu\Lb p_{1,T},
 k_T\Rb\,\Gamma_\mu\Lb p_{2,T}, k_T\Rb
\eeq
with integration over $k_T$. Since this function does not depend on $Q_T$, 
we are
 not interested in its exact structure.  The only function which determines the
 $Q_T$, is the triple Pomeron vertex (see Ref.\cite{GOLE}).
However, we recall that  in   inclusive  production, the contributions
 of  the BFKL Pomerons with rapidities $Y - y_1(y_2)$ and $y_1(y_2) - Y'$ vanish
 in deep saturation region, as they are proportional to $\nabla^2 N\Lb r,
 \dots\Rb$, ( where $r$ denotes the dipoles size \cite{KOTUINC,KOLEB}),
  and $N \to
 1$ in the saturation region. This means that the contributions of 
these Pomerons
 have maximum at $z \to 0$. Therefore, we can use the solutions of \eq{EQ6} to
 estimate the value of the cross section. 

To specify the $Q_T$ dependence, we need to find which values of $z'$( or 
$Y'$)  
contribute to the integral. Plugging in $T_A$ from \eq{EQ7}, we can take the
 integral over $z'$  resulting in the following expression
\beq \label{M3}
\frac{d^2 \sigma^{\rm interference\,diagram}}{d y_1 \,d y_2\,d^2 p_{1,T}\,d^2 p_{2,T}}\,\,\propto\,\, e^{2 \Lb 1 - \gamma_{cr}\Rb \, z } \,\frac{ 1}{\Lb 1\,+\, \phi_0\, \Lb e^{2 \Lb 1 - \gamma_{cr}\Rb\, z} - 1\Rb\Rb^2}\Lb \frac{ (1-\phi_0)}{1 - \gamma_{cr}}  \,+\, \phi_0 \,z_1 \Rb
\eeq
The two terms in \eq{M3} stem from different region of integration over 
$z'$. The
 first one  originates from  $z' \to 0$ or $Y' \propto 1/\bas$.  The second
 term comes from the region of integration in the entire kinematic region. The
 typical saturation momentum for such an integration is equal to
 $\bar{Q}^2_{s,A}\,=\,\sqrt{Q^2_{s,A}\Lb Y_0\Rb\,Q^2_{s,A}\Lb y_1\approx
 y_2 \Rb}$.

The dependence on $Q_T$  only comes from the triple Pomeron vertex.
 Since $G_A \propto S_A\Lb b \Rb$, the typical $Q_T$ along two upper
 BFKL Pomerons are equal to zero $Q_T \sim 1/R_A \ll 1/Q_s$, and the 
dependence on azimuthal angle $\varphi$ stems from $p^2_{12,T} \, =
 \,4 p^2_T \,\sin\Lb \varphi/2\Rb$. Finally, the general formula for
 the angular correlations has the form

\begin{figure}
 \includegraphics[width=10cm]{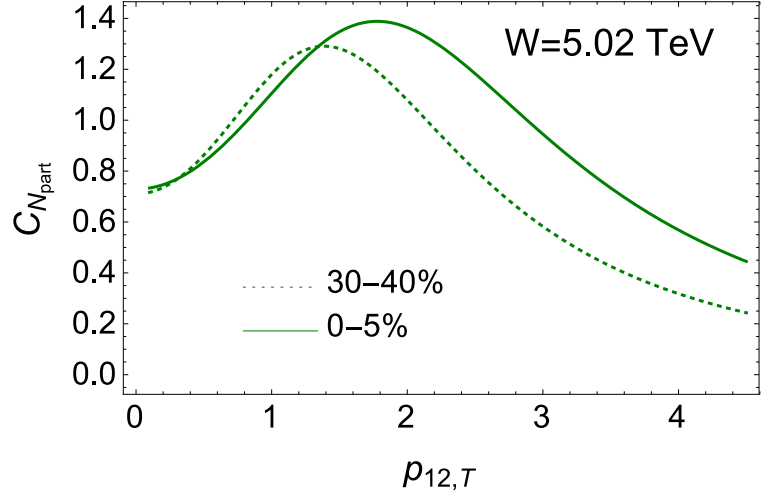}%
   \caption{ The correlation function $C_{N_{part}}\Lb p_{12,T}\Rb$ at different
 centralities: 0-5\%   and  30-40\% , versus $p_{12,T}$  }
\label{Cp}
   \end{figure}

\bea \label{M4}
\frac{d^2 \sigma}{d y_1 \,d y_2\,d^2 p_{1,T}\,d^2 p_{2,T}}\,\,\,&\propto&\, \,\,\Bigg(\frac{ (1-\phi_0)}{1 - \gamma_{cr}}\,\Gamma_{3 \pom}\Lb Q_T=0,Q_{s,A}\Lb Y_0\Rb\Rb   \,\,+\, \phi_0 \,z_1\,\Gamma_{3 \pom}\Lb Q_T =0 ,\bar{Q}_{s,A}\Lb y_1 \approx\,y_2\Rb\Rb\Bigg) \\
 \,\,&+&\,\,\frac{1}{N^2_c - 1}\,\Bigg(\frac{ (1-\phi_0)}{1 - \gamma_{cr}}\,\Gamma_{3 \pom}\Lb Q_T=p_{12,T},Q_{s,A}\Lb Y_0\Rb \Rb  \,\,+\, \phi_0 \,z_1\,\Gamma_{3 \pom}\Lb Q_T =p_{12,T} ,\bar{Q}_{s,A}\Lb y_1 \approx\,y_2\Rb\Rb\Bigg)\nn
\eea
\begin{figure}[!tbp]
  \centering
  \begin{minipage}[b]{0.45\textwidth}
    \includegraphics[width=\textwidth]{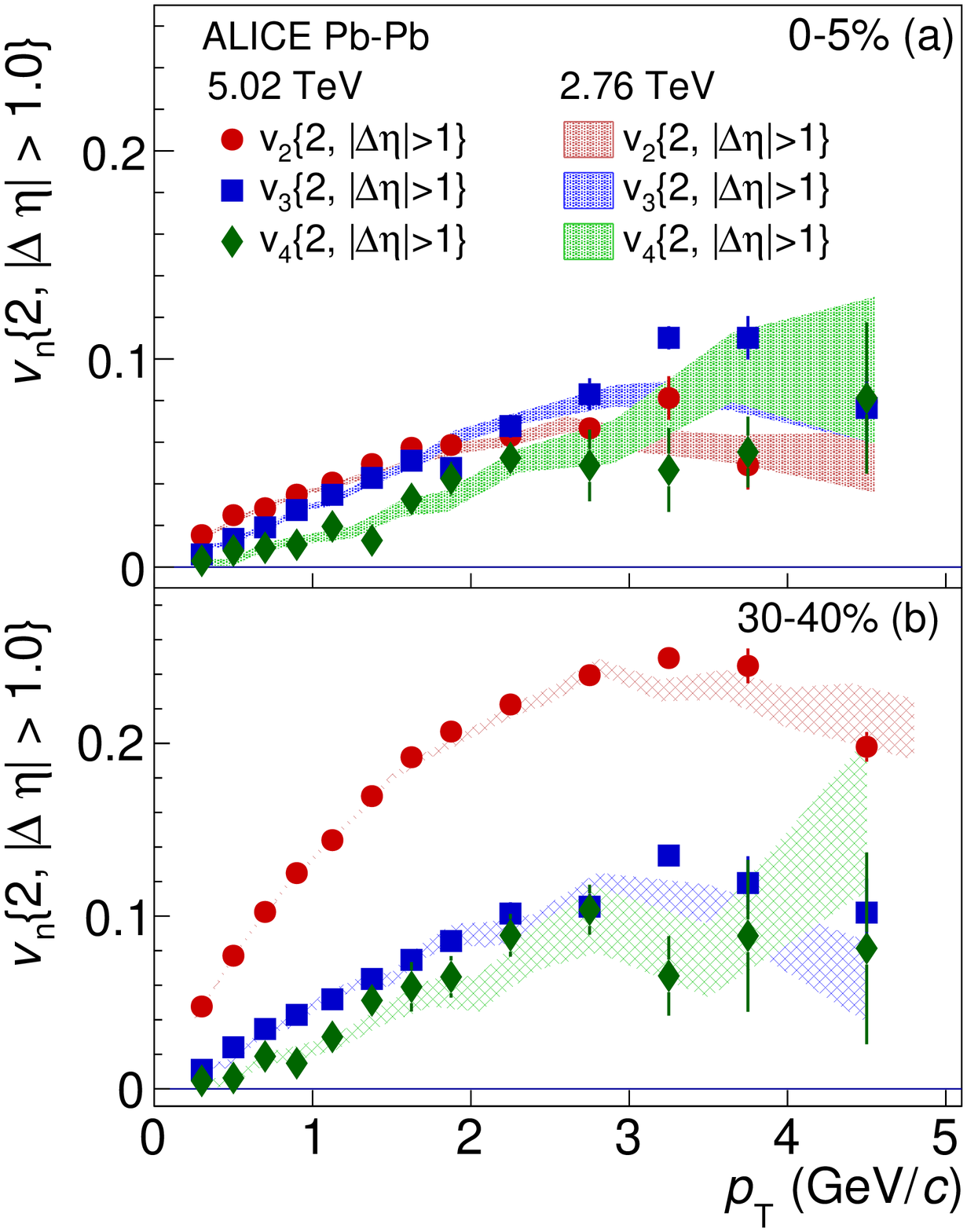}
    \caption{Experimental data for $v_n$ versus $p_T$\protect\cite{ALICEAA}
 at two different centralities: 0-5\%  in the upper figure and 30-40\% in
 the lower one.
  }
    \label{vnexp}
  \end{minipage}
  \hspace{0.9cm}
  \begin{minipage}[b]{0.39\textwidth}
    \begin{tabular}{c}
  \includegraphics[width=\textwidth ]{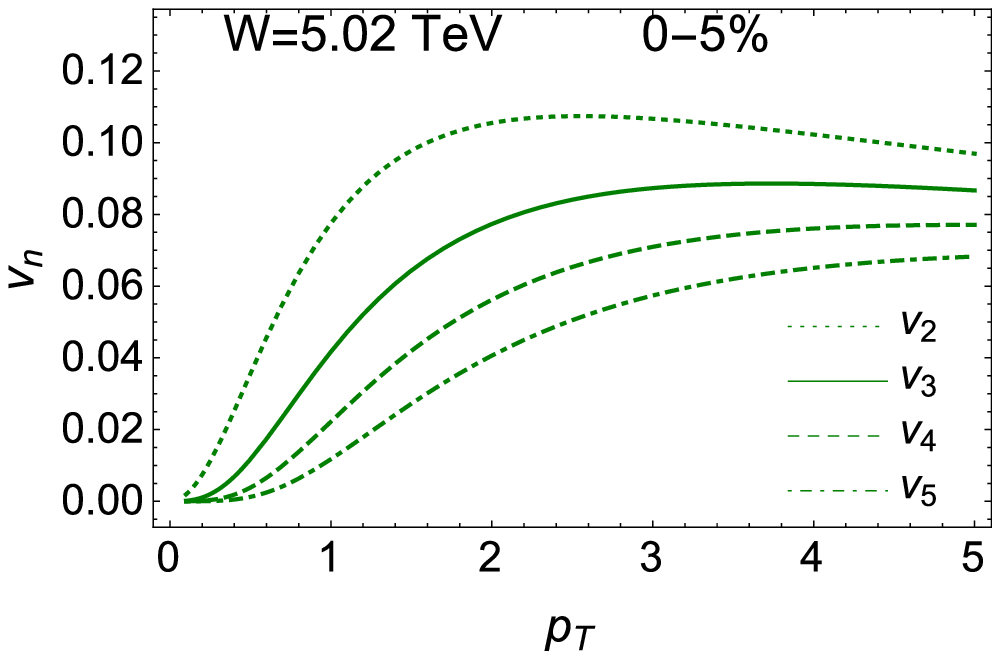}\\
   \includegraphics[width=\textwidth]{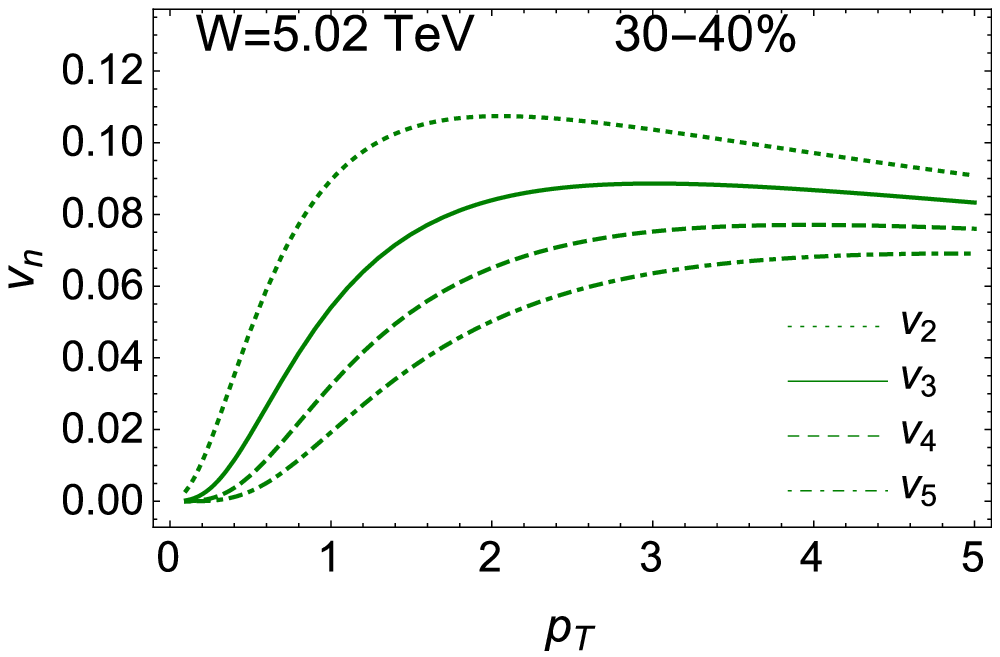} \\ 
\end{tabular}    
    \caption{Our model for $v_n$ versus $p_T$ for different
 centralities: 0-5\% in the upper  and 30-40\% in the lower figures.}
    \label{vnvspt}
  \end{minipage}
\end{figure}

The triple Pomeron vertex  has been calculated in Ref.\cite{GOLE}, and at
 large $Q_T$ it  has the form (see Eq.(45) and Eq.(A12))
\beq \label{M5}
\Gamma_{3 \pom}\Lb Q_T,Q_{s,A}\Rb\,\,\xrightarrow{Q_T 
 \,\gg\,Q_{s,A}} \,\,\Bigg(\frac{1}{\Lb \vec{k}_T - \h \vec{Q}_T\Rb^{2 \gamma_{cr}} \Lb Q^2_T\Rb^{1 - 2 \gamma_{cr}}}\Bigg)^2\,\,\,\xrightarrow{Q_T 
 \,\gg\,k_T \approx\,Q_{s,A}}\,\,\frac{1}{\Lb Q^2_T\Rb^{2 (1 - \gamma_{cr})}}
 \eeq
where $k_T$ denotes the momentum inside of the triple Pomeron vertex, which is of 
the 
order of the typical saturation momentum of the lower BFKL Pomeron in \fig{incl}.
To  specify dependence of the triple Pomeron vertex, we recall that 
at large 
impact
 parameters, the scattering amplitude should decreases exponentially\cite{FROI}
.
 Bearing this in mind we suggest that
\beq \label{M6}
\Gamma_{3 \pom}\Lb Q_T,Q_{s,A}\Rb\,\,=\,\,\Bigg( \frac{Q^2_s}{Q^2_T \,\,+
\,\,Q^2_s}\bigg)^{2 ( 1 - \gamma_{cr})}
 \eeq
 which reproduces \eq{M5} at large $Q_T$, and has the exponential decrease 
at large $b$.

\begin{figure}
 \includegraphics[width=10cm]{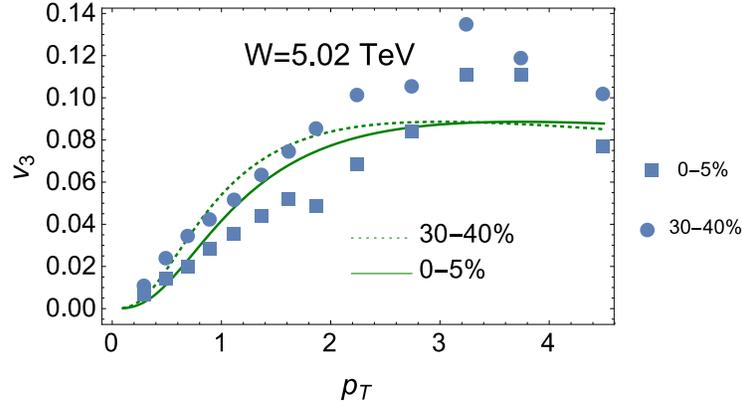}%
   \caption{ Comparison of the estimates from our model for
 $v_3$, with the experimental
 data of ALICE collaboration\cite{ALICEAA}. 
}
\label{v3exp}
   \end{figure}


Plugging \eq{M6} into \eq{M4} we obtain the correlation function in the form
\beq \label{M7}
C_A\Lb p_{12,T}\Rb\,\,=\,\,\frac{1}{N^2_c - 1}\frac{\Bigg(\frac{ (1-\phi_0)}{1 - \gamma_{cr}}\,\Gamma_{3 \pom}\Lb Q_T=p_{12,T},Q_{s,A}\Lb Y_0\Rb \Rb  \,\,+\, \phi_0 \,z_1\,\Gamma_{3 \pom}\Lb Q_T =p_{12,T} ,\,\bar{Q}_{s,A}\Lb y_1 \approx\,y_2\Rb\Rb\Bigg)}{\Bigg(\frac{ (1-\phi_0)}{1 - \gamma_{cr}}\,\Gamma_{3 \pom}\Lb Q_T=0,Q_{s,A}\Lb Y_0\Rb\Rb   \,\,+\, \phi_0 \,z_1\,\Gamma_{3 \pom}\Lb Q_T =0 ,\bar{Q}_{s,A}\Lb y_1 \approx\,y_2\Rb\Rb\Bigg)}
\eeq

The multiplicity dependence stems from \eq{M7}, where we replace
 $Q_{S,A}$ by the value of the saturation momentum, which corresponds
 to the given number of participants, this in the spirit  of the
 KLN approach\cite{KLN5,KLN6}.
In \fig{Cp} the correlation functions are shown for W = 5.02\,TeV, and for
 the choice $Y_0 = \ln\Lb W_0/m\Rb$ with $W = 130 \,GeV$ and $m = 1\,GeV$.
   This function has an essential dependence on $N_{\rm part}$,
 or on centrality.

$v_n$ can be calculated  for $|p_{1,T}|\,=\,|p_{2,T}|$ as
\beq \label{M8}
v_{n} \,\,=\,\,\Bigg( \int d \varphi \cos\Lb n \varphi\Rb \,C_{N_{\rm part}}
\Lb 2 p_{T}\sin\Lb \varphi/2\Rb\Rb \Bigg{/} \Lb 2 \pi\,\,+\,\,\int d \varphi\, C_{N_{\rm part}}\Lb 2 p_{T}\sin\Lb \varphi/2\Rb\Rb\Rb\Bigg)^\h
\eeq


\subsubsection{The Choice of  parameters.}

The  formulae of \eq{M7} and \eq{M8} depend only on the value of
 the saturation momentum, and consequently, it depends on rapidity, 
and $N_{part}$.
 We follow 
the KLN-approach\cite{KLN1,KLN4,KLN5}  in finding  these dependences.
  We assume that
\beq \label{PC1}
Q^2_{s}\Lb Y; N_{\rm part}\Rb \,\,=\,\  \frac{\rho_{\rm part}}{2}\,Q^2_0
\,e^{\lambda \Lb  Y- Y_0\Rb};
\eeq
The value of $Q_0$ we fix from the gold-gold scattering  at $W=130 \,GeV$ and
 for centrality $0 - 5\%$  $Q^2_s\Lb Y=Y_0\Rb\,=\,2.02 \,GeV^2$. $Y = \ln \Lb
 W/W_0\Rb$ and $ Y - Y_0 = \ln\Lb W/130\Rb$. $\rho_{\rm part}$ have been
 calculated in Ref.\cite{KLN1} for the LHC energies, and in Ref.\cite{KLN5}
 for $W_0 = 130\,GeV$. The choice $Y_0= \ln\Lb W_0/m\Rb$ in  \eq{M7} is not
 theoretically determined, note that the value of typical
 $\Delta Y' $ in the integral over $Y'$,  is about $\Delta Y \sim 1/\bas$,
 and for $\bas = 0.2$, this results in a  value which is close to 
the chosen
 $Y_0$. Finally, we take $\lambda = 0.25$ as it is done in  
 Refs.\cite{KLN1,KLN2,KLN3,KLN4,KLN5}.


\subsubsection{Comparison with the experimental data.}


Using the parameters, discussed above, we evaluate the correlation function
 (see \fig{Cp}, and the values of $v_n$ which are plotted in 
\fig{vnexp}, \fig{vnvspt}
 and \fig{v3exp}). First, we  note that the correlation function  
depends  strongly
  on the centrality, leading to a correlation length which  increases for
 large centralities.

However, $v_n$ show only mild dependence on centralities (compare \fig{vnexp}
 and \fig{vnvspt} ). Such a behaviour  at first sight is in disaccord with the
 experimental data.  $v_2$ turns out to be smaller that the experimental values
 for both centralities. On the other hand, the value for $v_2$,
 as well as for other
 even $n$, is not very  decisive, since in QCD  there are  many other 
sources 
of $v_n$ 
with
 even $n$,  beside Bose-Einstein correlations. However, we have not 
found
 the other sources for $v_n$ with even $n$. \fig{v3exp} presents
 our estimate for
 $v_3$ together with the experimental data.  We see that our predictions for
 $v_3$ describe the experimental data fairly well. Not extremely well, but the
 model that we develop here, is very simple. These estimates encourage us to
 develop a  more complete description of $v_n$ for  even $n$, with different
 multiplicities,  based on the Bose-Einstein correlations.

\section{Conclusions}
  
 We  summarize the main results of this paper. The main goal of 
this paper
 is to investigate the dependence of  Bose-Einnstein correlations on the
 multiplicity of the events. We view these correlations as the major source
 of the azimuthal angle correlations, and the only known origin of $v_n$ with
 odd $n$ in the framework of the Color Glass Condensate. Indeed, the
 correlation of identical gluons  produces the correlation function 
that
 depends on $|\vec{p}_{1,T} - \vec{p}_{2,T}|$   which gives $v_n$
 with odd $n$. However, in Refs.\cite{KOWE,KOLUCOR} it was noted, that
 the diffractive central production of two different gluons in the colorless
 state leads to dependence on
$|\vec{p}_{1,T} + \vec{p}_{2,T}|$. If these two sources have the same
 strength,
 the totally inclusive experiments without any selection on multiplicities, will
 give $v_n=0$ for odd $n$. In this paper, we showed in the Leading Log(1/x)
 Approximation of perturbative QCD, the amplitude of two gluon
 exclusive production turns out to be equal to  the interference
 diagram, that is the source of 
the
 Bose-Einstein correlation, in accord with Refs.\cite{KOWE,KOLUCOR}. However, 
 the emission of soft gluons for the central exclusive production
 in the Double Log Approximation  of perturbative QCD, leads to the 
Sudakov
 form factor which  suppress this contribution.  Therefore,  the Bose-Einstein
 correlations
 prevail leading to $v_n \,\neq\,0$ for odd $n$.  It
 should be stressed, that without this suppression,
 the measurement of an event with given multiplicity,
 yields $v_{n,n}\,<\,0 $ for odd $n$.

We demonstrated that the Bose-Einstein correlation function does not depend on 
the number of produced parton showers for hadron-nucleus and nucleus-nucleus
 collisions, but for hadron-hadron collisions such dependence turns out to be
considerable.

Finally, we developed a simple KLN-type model to describe the  Bose-Einstein
 correlation in one parton cascade, as a function of centralities.
  The predicted
 dependence reflects the main features of the observed data, reproduces
the value of $v_n$ with odd $n$, but, much work is still needed
 to develop a 
more
 complete approach. This paper encourages us to search for such an approach. 

We view this paper as an argument that the description of $v_n$ is
 possible due to
 interactions in the initial state, and that these interactions 
 should not 
 be neglected.


  {\bf Acknowledgements} 
   We thank our colleagues at Tel Aviv University and UTFSM for
 encouraging discussions. Our special thanks go to    
Carlos Cantreras, Alex Kovner and Michael  Lublinsky for
 elucidating discussions on the
 subject of this paper. 
 
 This research was supported by the BSF grant   2012124, by    Proyecto Basal FB 0821(Chile) ,  Fondecyt (Chile) grant   1140842, and by   CONICYT grant PIA ACT1406.  
 ~

    \appendix
{}

\section{Integration over longitudinal momenta}

\begin{figure}[ht]
 \includegraphics[width=14cm]{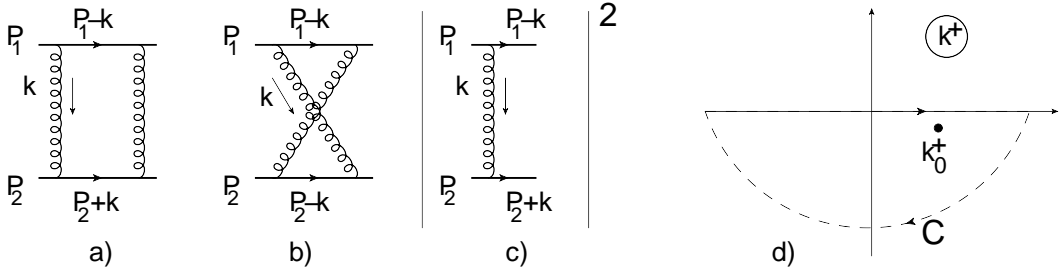}
   \protect\caption{Born Approximation of pQCD: longitudinal
 momenta integration. \fig{a1}-a and \fig{a1}-b  are the diagrams for
 the scattering amplitude at high energy in the $\bas^2$ order of pQCD.
 \fig{a1}-c is the cross section for two quarks production (cut Pomeron).
 \fig{a1}-d shows the contour of integration over $k^+$. Helical lines 
denote
 gluons, the solid lines indicate quarks. }
\label{a1}
   \end{figure}

In this appendix we recall the calculation that results in \eq{UNIT1}.
 For simplicity  we restrict ourselves to calculate both the scattering
 amplitude at high energies ( Pomeron, see \fig{a1}-a and \fig{a1}-b) 
 and the contribution of the inelastic processes (cut Pomeron, $G\Lb s
 \Rb$ in \eq{UNIT1}, see \fig{a1}-c), in the Born approximation of pQCD.
  $G\Lb s,t=0\Rb$ takes the following form (see \fig{a1}-c)
\bea \label{A101}
G\Lb s,t=0\Rb\,&=&\,\,g^4 \,{\cal C} \,4 s^2 \, \int \frac{ d k^+ d  k_-\,d^2 k_T}{(2\,pi)^4\, i }\frac{1}{\Lb  k^+ k_{-}\,-\,k^2_T - i \epsilon\Rb^2}\,2 \pi \delta\Lb (P_1 -?k)^2\Rb\,2 \pi \,\delta\Lb (P_2 +k)^2\Rb\,\nn\\
&=&\,16\, {\cal C} \,\as^2 \,s^2\, \int d k^+ d  k_-\,d^2 k_T\,\frac{1}{\Lb  k^+ k_{-}\,-\,k^2_T - i \epsilon\Rb^2}\,\delta\Lb - P_1{^+}\,k_{-} -\, k^2_T\Rb\,\delta\Lb P_{2,-}\,k^{+} -\, k^2_T\Rb\nn\\
&=& 16\, {\cal C} \,\as^2 \,s\,\int \frac{d^2k_T}{k^4_T}
\eea
 In \eq{A101} ${\cal C}$ is the colour coefficient which is the same for
 all diagrams, factor $4 s^2$  ($s = \Lb P_1 + P_2\Rb^2 = 2 P_{1,\mu}
\,P^{\mu}_2$ at high energy) stems from the summation over polarization
 of the $t$-channel gluon of the gluon current of quarks $ 2 P_{1,\mu}$
 ($2 P_{2,\mu}$). $\as = g^2/4 \pi$.   Integrating the $\delta$-functions, 
one
 can see that $k^+\,k_{-} \,\,\ll\,k^2_T$.

The scattering amplitude is equal to
\bea \label{A10}
A\Lb s, t=0\Rb\,\,&=&\,\,g^4 \, {\cal C}\, 4 \,s^2\, \int \frac{ d k^+ d  k_-\,d^2 k_T}{(2\,pi)^4\, i } \frac{1}{\Lb  k^+ k_{-}\,-\,k^2_T - i \epsilon\Rb^2}\,\frac{1}{- P_1{^+}\,k_{-} -\, k^2_T \,-\,i\,\epsilon}\nn\\
 &\times& \Bigg( \underbrace{\frac{1}{P_{2,-}\,k^+ - k^2_T \,-\,i \,\epsilon}}_{\fig{a1}-a}\,+\,\underbrace{\frac{1}{- P_{2,-}\,k^+ - k^2_T \,-\,i \,\epsilon}}_{\fig{a1}-b}\Bigg)
\eea

For $k^+ \,>\,0$ we can take the integral over the pole: $ k^0_-\,=
\,\frac{-k^2_T- i \epsilon}{P^+_1}$ closing around this pole, the  contour  
of 
integration in lower semi-plane  in complex $k_{-}$  plane, since the integral over large circle decreases at large $k_-$.  The other pole $k^1_-\,= \,\frac{k^2_T +  i \epsilon}{k^+}$ is located in the upper semi-plane.  For  $ k^+  \,<\,0$ all singularities are situated in lower semi-plane leading to vanishing of the integral. Bearing this in mind we reduce \eq{A10} to the following expression:
\bea \label{A11}
A\Lb s, t=0\Rb\,\,&=&\,\,\frac{8\,\as^2 }{ \pi} \,{\cal C}\, s^2\, \int^\infty_0 \,d\, k^+ d^2 k_T\, \frac{1}{k^4_T}\,\frac{1}{\Lb - P_1^+ \Rb} \Bigg( \frac{1}{P_{2,-}\,k^+ - k^2_T \,-\,i \,\epsilon}\,+\,\frac{1}{- P_{2,-}\,k^+ - k^2_T \,-\,i \,\epsilon}\Bigg)\nn\\
& \,=\,&\,\,\frac{8\,\as^2 }{ \pi} \,{\cal C}\, s^2\, \int^\infty_{-\infty}  \,d\, k^+ d^2 k_T\, \frac{1}{k^4_T}\,\frac{1}{\Lb - P_1^+ \Rb}  \frac{1}{P_{2,-}\,k^+ - k^2_T \,-\,i \,\epsilon}
\eea
Taking the integral over $k^+$ using contour $C$ in \fig{a1}-d,  and 
taking 
into account that the integral over  a large circle is equal to $ i\,\pi$ 
we obtain
\beq \label{A12}
A\Lb s, t=0\Rb\,\,=\,\,\,i\,8\,\as^2 \,{\cal  C}\, s\,\int \frac{d^2 k_T}{k^4_T}
\eeq

The diagram \fig{a1}-c gives the same contribution as the imaginary
 part of diagram of \fig{a1}-a,  multiplied by factor 2, since in this
 diagram we have $2 \pi 
\delta\Lb P_{2,-}\,k^+ - k^2_T\Rb$. Therefore, we obtain that $2\,
 \mbox{Im}\, A\Lb s,t=0\Rb = G\Lb \fig{a1}-c\Rb$ which proves
 \eq{UNIT1} in Born approximation of pQCD. 

\begin{figure}[ht]
 \includegraphics[width=10cm]{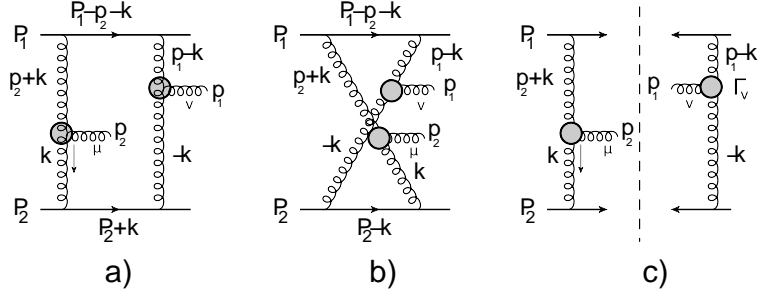}
   \protect\caption{Born Approximation of pQCD: longitudinal momenta
 integration. \fig{a2}-a and \fig{a2}-b are the diagrams for the amplitude
 for  the production of two gluons with momenta $p_{1}$ and $p_{2}$, 
 in $\bas^3$ order of pQCD. \fig{a2}-c is the cross section for two
 quarks and two gluons  production (cut Pomeron).  Helical lines denote
 gluons, the solid lines indicate quarks. The blobs denote the Lipatov
 vertices for gluon production ($\Gamma_\nu$).}
\label{a2}
   \end{figure}

  For the amplitude of the two gluon production (see \fig{a2}-a and
 \fig{a2}-b) as well as for the cross section of the one gluon production
 which is shown in \fig{a2}-c, we have the following hierarchy of the
 longitudinal momenta:
  \beq \label{A14}
  P^+_1  \,\gg\,p^+_1 \sim p^+_2\,\gg\,k^+;~~~~~~~~~~ P_{2,-} 
 \,\gg\,p_{1,-} \sim p_{2,-}\,\gg\,k_{-};
  \eeq
    assuming that both gluons are produced with almost equal rapidities
 ($y_1 \approx y_2$) in the central rapidity region ($y_1 \approx y_2 \ll
 1$) in c.m.f. 
    
    Using \eq{A14} we can reduce the amplitude to the following expresion:
    \bea\label{A15}
  &&  A\Lb \fig{a2}-a + \fig{a2}-b\Rb\,\,=\,\\
  &&  \,32\,\pi\,\as^3  \,{\cal C}\, s^2\, \int^\infty_0 \,d\, k^+ d^2 k_T\, \frac{1}{k^4_T}\,
    \frac{\Gamma_\mu\Lb p_{2,T},k_T\Rb}{\Lb \vec{p}_{2,T} +\vec{k}_T\Rb^2}\,\frac{\Gamma_\nu\Lb p_{1,T},k_T\Rb}{\Lb \vec{p}_{1,T}-\vec{k}_T\Rb^2}\,\,\frac{1}{\Lb - P_1^+ \Rb} \Bigg( \frac{1}{P_{2,-}\,k^+ - k^2_T \,-\,i \,\epsilon}\,+\,\frac{1}{- P_{2,-}\,k^+ - k^2_T \,-\,i \,\epsilon}\Bigg)\nn\\
&& \,=\,\,\,32\,\,\as^3  \,{\cal C}\, s^2\, \int^\infty_{-\infty}  \,d\, k^+ d^2 k_T\, \frac{1}{k^4_T}\,\frac{\Gamma_\mu\Lb p_{2,T},k_T\Rb}{\Lb \vec{p}_{2,T} +\vec{k}_T\Rb^2}\,\frac{\Gamma_\nu\Lb p_{1,T},k_T\Rb}{\Lb \vec{p}_{1,T}-\vec{k}_T\Rb^2}\frac{1}{\Lb - P_1^+ \Rb}  \frac{1}{P_{2,-}\,k^+ - k^2_T \,-\,i \,\epsilon} \nn\\
&& \,=\,32\,\pi \,i\,\as^3  \,{\cal C}\, s \int \frac{\Gamma_\mu\Lb p_{2,T},k_T\Rb\,\Gamma_\nu\Lb p_{1,T},k_T\Rb\,d^2 k_T}{k^4_T\, \Lb \vec{p}_{2,T} +\vec{k}_T\Rb^2\,\Lb \vec{p}_{1,T}-\vec{k}_T\Rb^2}\nn
\eea
In \eq{A15} we use the same contour of integration over $k^+$ (see 
\fig{a1}-d) as calculating the elastic amplitude (see \eq{A11}). 
 The Lipatov vertices $\Gamma_\mu$ for the gluon emission depend
 only on transverse momenta, and do not influence  the integration
 over longitudinal momenta.

The cross section of \fig{a2}-c differs from the amplitude by factor 2,
 which has the same origin as has been discussed above (see \eq{A101}.

     
   
   \section{Vertices for two gluon production in the central rapidity region}
   
     In this appendix we clarify why diagrams \fig{vG}-a and \fig{vG}-b are equal.  The vertex in the diagram of \fig{vG}-a has the form
     
     \beq \label{VERA}
  V\Lb \fig{vG}-a\Rb\,\,=\,\,
      \frac{\Gamma_\mu\Lb \vec{k}_T, \vec{p}_{1,T}\Rb\, \Gamma_{\nu}\Lb - \vec{k}_T, \vec{p}_{2,T}\Rb}{k^2_T\, \Lb \vec{k}_T - \vec{p}_{2,T}\Rb^2} 
      \eeq
      while for \fig{vG}-b it can be written as
      \beq \label{VERB}
      V\Lb \fig{vG}-b\Rb\,\,=\,\,
      \frac{\Gamma_\mu\Lb \vec{k}_T,\vec{p}_{1,T}\Rb\, \Gamma_{\nu}\Lb \vec{k}_T - \vec{p}_{1,T} , \vec{p}_{2,T}\Rb}{k^2_T\, \Lb \vec{k}_T - \vec{p}_{2,T}\Rb^2}     
      \eeq  
      We need to calculate these vertices for $\vec{p}_{1,T} \,=\,-\,\vec{p}_{2,T}$, since $|\vec{p}_{1,T} \,+\,\,\vec{p}_{2,T} |\propto 1/R_D \ll 1/R_N$.
      
      The vertices $\Gamma_\mu$ has the following expressions:
      \beq \label{B3}
      \Gamma_\mu\Lb \vec{k}_T,\vec{p}_{1,T}\Rb = \frac{1}{p^2_{1,T}}\Lb k^2
 \vec{p}_{1,T} - p^2_{1,T} \,\vec{k}_T\Rb; \\
       ~~
          \Gamma_{\nu}\Lb \vec{k}_T - \vec{p}_{1,T} , \vec{p}_{2,T}\Rb =\frac{1}{p^2_{2,T} }\Lb ( \vec{k} - \vec{p}_{1,T})^2 \vec{p}_{2,T} - p^2_{2,T} \Lb \vec{k}_T -\vec{p}_{1,T}\Rb\Rb;
          \eeq

               We need to convolute these vertices with
 $   \Gamma_\mu\Lb \vec{l}_T,\vec{p}_{1,T}\Rb  $ and $  
   \Gamma_{\nu}\Lb \vec{l}_T - \vec{p}_{1,T} , \vec{p}_{2,T}\Rb$
  for the different Pomerons, where the integration  is over 
$l_T$. 
  In such convolution      
the terms that are proportional to $\vec{p}_{1,T} \cdot \vec{k}_T$ or to
      $\vec{p}_{1,T} \cdot \vec{l}_T$     vanish due to angular 
integrations.
      Only the term which is proportional to $\Lb \vec{k}_T \cdot
 \vec{l}_T\Rb^2$ survives and  yelds $\h l^2_T k^2_T$.
   It is easy to see that this term is the same in both vertices
 of \eq{VERA} and \eq{VERB}.    Now we need to compare
   \bea \label{B4}        
\Gamma_\mu\Lb \vec{k}_T,\vec{p}_{T}\Rb\cdot \vec{p}_T& = &\frac{1}{p^2_{T}}\Lb
 k^2 \vec{p}_{T} - p^2_{T} \,\vec{k}_T\Rb\cdot \vec{p}_T 
 \eea
with 
\bea \label{B41}
\Gamma_{\nu}\Lb \vec{k}_T - \vec{p}_{T} , \vec{p}_{T}\Rb =\frac{1}{p^2_{T}
 }\Lb ( \vec{k} - \vec{p}_{T})^2 \vec{p}_{T} + p^2_{T} \Lb \vec{k}_T
 -\vec{p}_{T}\Rb\Rb\cdot \vec{p}_T;
\eea
where we denote $\vec{p}_{1,T} = \vec{p}_T = - \vec{p}_{2,T}$.

The direct calculations  gives the same expression for both terms: 
\beq \label{B5}
\h p^2_T\Lb \Lb \vec{k}_T - \vec{p}_T\Rb^2 + k^2_T - p^2_T\Rb
\eeq

Therefore,  both diagrams give the same contribution.

 \end{document}